\documentclass[11pt]{article}
\usepackage{setspace}

\usepackage{titling}
\usepackage{booktabs}
\usepackage{array}
\usepackage{tikz}
\usetikzlibrary{arrows.meta}
\usepackage{natbib}
\usepackage{calc}
\usepackage{subcaption}
\usepackage[a4paper, margin=1in]{geometry}
\usepackage{rotating}
\usepackage{booktabs}
\usepackage{mathrsfs}
\usepackage{fancyhdr}
\usepackage{amsfonts}
\usepackage{amstext}
\usepackage[T1]{fontenc}     % Font encoding
\usepackage{lmodern}         % Latin Modern fonts (enhanced Computer Modern)
\usepackage{amsmath,amsthm} 
\usepackage{amssymb}
\usepackage{algorithm}
\usepackage{algorithmic}
\usepackage{adjustbox}
\usepackage{hyperref}
\hypersetup{ hidelinks }
\hypersetup{breaklinks=true}
\usepackage{bookmark}
\usepackage{setspace}
\usepackage{longtable}
\usepackage{float}
\usepackage{threeparttable}
\usepackage{fullpage}
\usepackage{times}
\usepackage[utf8]{inputenc}
\usepackage{units}

\usepackage{nicefrac}
\usepackage{color}
\usepackage{tabularx}
\usepackage{pdflscape}
\usepackage{xcolor}

\newtheorem{theorem}{Theorem}
\newtheorem{assumption}{Assumption}
\newtheorem{corollary}{Corollary}
\newtheorem{lemma}{Lemma}
\newtheorem{remark}{Remark}

\newcommand{\ii}{{\rm i}}

\pretitle{\begin{center}\large\bfseries}
\posttitle{\end{center}}
\preauthor{\begin{center}
\large \lineskip 0.5em%
\begin{tabular}[t]{c}}
\postauthor{\end{tabular}\par\end{center}}
\predate{\begin{center}\large}
\postdate{\par\end{center}\vskip 0.5em}

\newcommand{\HRule}{\rule{\linewidth}{0.5mm}}
\newcommand{\BRule}{\rule{\linewidth}{0.25mm}}
\title{\HRule \\[0.1mm]
Finite-Sample Distortion in Kernel Specification Tests: A Perturbation Analysis of Empirical Directional Components
\\[0.1mm]
\BRule}

\author{Rui Cui\thanks{cuirui.econ@outlook.com} \\ \small Guangzhou Nanfang College
\and 
Yuhao Li\thanks{yuhao.li@xjtlu.edu.cn} \\ \small Xi'an Jiaotong-Liverpool University
\and 
Xiaojun Song\thanks{sxj@gsm.pku.edu.cn} \\ \small Peking University
}

\date{} % No date

\begin{document}
%\onehalfspacing
\maketitle
%\onehalfspacing
\begin{abstract}
    This paper provides a new theoretical lens for understanding the finite-sample performance of kernel-based specification tests, such as the Kernel Conditional Moment (KCM) test. Rather than introducing a fundamentally new test, we isolate and rigorously analyze the finite-sample distortion arising from the discrepancy between the empirical and population eigenspaces of the kernel operator. Using perturbation theory for compact operators, we show that the estimation error in directional components is governed by local eigengaps: components associated with small eigenvalues are highly unstable and contribute mostly noise rather than signal under fixed alternatives. Although this error vanishes asymptotically under the null, it can substantially degrade power in finite samples. This insight explains why the effective power of omnibus kernel tests is often concentrated in a low-dimensional subspace. We illustrate how truncating unstable high-frequency components--a natural consequence of our analysis--can improve finite-sample performance, but emphasize that the core contribution is the diagnostic understanding of \textit{why} and \textit{when} such instability occurs. The analysis is largely non-asymptotic and applies broadly to reproducing kernel Hilbert space based inference.
\end{abstract}

\vspace{1em} % Add some vertical space between the abstract and the keywords
\noindent \textbf{Keywords:} Convergence of eigenspaces; Perturbation Theory; Reproducing kernel Hilbert space; Specification test; Truncation

\pagenumbering{arabic} \setcounter{page}{1} \setcounter{footnote}{0}

\section{Introduction}
\label{sec:intro}

We consider the following general parametric regression model:
\begin{equation*}
    Y = \mathcal{M}_{\theta_0}(X) + \varepsilon_{\theta_0},
    \label{eq:cmr}
\end{equation*}
where $\mathcal{M}_{\theta_0}(X)$ is a parametric function indexed by the unknown parameter vector $\theta_0 \in \Theta \subset \mathbb{R}^d$, $X \in \mathcal{X} \subset \mathbb{R}^d$ denotes the covariate, $Y \in \mathcal{Y} \subset \mathbb{R}$ is the response variable, and $\varepsilon_{\theta_0}$ represents the model error. This framework encompasses a broad class of models, including linear and nonlinear regression, treatment effect models, and instrumental variables models. Under the null hypothesis of correct model specification, the error satisfies the conditional moment restriction:
$$ 
    \mathbb{E}(\varepsilon_{\theta_0} \mid X) = 0 \quad \text{a.s.}
$$

A standard strategy for testing model specification transforms this conditional moment restriction into an infinite collection of unconditional moment conditions over a rich class of test functions. Specifically,
$$ 
    \mathbb{E}(\varepsilon_{\theta_0} \mid X) = 0 \text{ a.s. } \iff \mathbb{E}\big(\varepsilon_{\theta_0} h(X, \cdot)\big) = \boldsymbol{0} \in \mathcal{H}, \quad \forall h \in \mathcal{H}, 
$$ 
where the expectation is interpreted in the Bochner sense. $\mathcal{H}$ is a class of measurable functions of $X$, see \cite{bierens1982consistent,bierens1997asymptotic,stute1997nonparametric,Stinchcombe_White_1998,delgado2006consistent,escanciano2006consistent,escanciano2024gaussian} for some examples. This approach, known as the \textit{Integrated Conditional Moment} (ICM) method, has been widely studied since its introduction. More recently, attention has turned to restricting $\mathcal{H}$ to a reproducing kernel Hilbert space (RKHS), giving rise to kernel-based specification tests:
$$ 
    \mathbb{E}(\varepsilon_{\theta_0} \mid X) = 0 \text{ a.s. } \iff \mathbb{E}\big(\varepsilon_{\theta_0} k(X, \cdot)\big) = \boldsymbol{0} \in \mathcal{H}_k,
$$ 
where $k(\cdot, \cdot)$ is a reproducing kernel associated with the RKHS $\mathcal{H}_k$. Exploiting the reproducing property of RKHS, one can show that the squared norm of this unconditional moment condition in $\mathcal{H}_k$ simplifies to:
$$ 
    \mathbb{E}\big(\varepsilon_{\theta_0} k(X, X') \varepsilon_{\theta_0}'\big) = 0,
$$
where $X'$ and $\varepsilon_{\theta_0}'$ are independent and identically distributed copies of $X$ and $\varepsilon_{\theta_0}$, respectively.

Building on this equivalence, a practical test statistic is constructed by replacing the unknown parameter $\theta_0$ with a consistent estimator $\hat{\theta}_n$ and substituting population expectations with their empirical counterparts. The resulting statistic is given by:
\begin{align*}
    \hat{T}_n & = \frac{1}{n} \boldsymbol{\varepsilon}_{\hat{\theta}}^\top (\boldsymbol{K}/n) \boldsymbol{\varepsilon}_{\hat{\theta}}  = \frac{1}{n} \sum_{j=1}^n \hat{\lambda}_j \left( \boldsymbol{U}_j^\top \boldsymbol{\varepsilon}_{\hat{\theta}_n} \right)^2,
\end{align*}
where $\boldsymbol{K}$ is the $n \times n$ kernel matrix with entries $\boldsymbol{K}_{\alpha \beta} = k(x_\alpha, x_\beta)$, and $\boldsymbol{\varepsilon}_{\hat{\theta}}$ denotes the vector of residuals evaluated at $\hat{\theta}_n$. The pairs $\{\hat{\lambda}_j, \boldsymbol{U}_j\}_{j=1}^n$ represent the eigenvalues and corresponding orthonormal eigenvectors of the scaled kernel matrix $\boldsymbol{K}/n$. To the best of our knowledge, \cite{muandet2020kernel} were the first to propose the \textit{Kernel Conditional Moment} (KCM) test based on this principle.

A substantial body of literature has examined the so-called parameter uncertainty effect--the distortion in inference caused by replacing the true but unknown parameter $\theta_0$ with its consistent estimator $\hat{\theta}_n$. In the classical ICM framework, foundational contributions include \cite{delgado2006consistent, delgado2008distribution, escanciano2006consistent}; more recently, \cite{escanciano2024gaussian} extended this analysis to the kernel-based setting.

Another important but less-explored source of distortion in the KCM test arises from the discrepancy between the empirical and population eigensystems. To understand this, consider that commonly used kernels--such as the Gaussian and Laplace kernels--are \textit{Mercer kernels}, which admit the following spectral decomposition:
$$ 
    k(s, t) = \sum_{j=1}^{\infty} \lambda_j \phi_j(s) \phi_j(t),
$$
where $\{\lambda_j\}_{j=1}^{\infty}$ are positive eigenvalues arranged in descending order ($\lambda_1 > \lambda_2 > \cdots > 0$), and $\{\phi_j(\cdot)\}_{j=1}^{\infty}$ are the corresponding orthonormal eigenfunctions of an integral operator (discussed in Section \ref{sec:RKHS_intro}). Using this representation, the squared norm of the population moment condition can be expressed as:
$$ 
    \mathbb{E}\big(\varepsilon_{\theta_0} k(X, X') \varepsilon_{\theta_0}'\big) = \sum_{j=1}^{\infty} \lambda_j \left( \mathbb{E}\big(\varepsilon_{\theta_0} \phi_j(X)\big) \right)^2.
$$

The eigensystem of the scaled kernel matrix $\boldsymbol{K}/n$ provides empirical estimates of the population eigensystem $\{\lambda_j, \phi_j\}_{j=1}^\infty$. Specifically, we have $| \hat{\lambda}_j - \lambda_j | = O_p(n^{-1/2})$ for all $j$; see Lemmas~\ref{lemma:upper_bound_operator} and~\ref{lemma:HS_bound} in the Appendix. For the eigenfunctions, we have $\sqrt{n}\boldsymbol{U}_j = \hat{\phi}_j(\boldsymbol{x})$ and $\| \hat{\phi}_j - \phi_j \|_{L_2(\mathbb{P})} = O_p(n^{-1/2})$ for all $j$; see Theorem~\ref{thm:upper_bound_eigenfunction} and Lemma~\ref{lemma:analytic_perturbation}. Here, $\hat{\phi}_j(\boldsymbol{x})$ denotes the $j$-th estimated eigenfunction evaluated at the observed covariates $\boldsymbol{x} = (x_1,\ldots,x_n)^\top$, $L_2(\mathbb{P})$ denotes the space of square-integrable functions with respect to the probability measure $\mathbb{P}$. Consequently, the empirical KCM test statistic can be expressed as:
$$ 
    \hat{T}_n =  \sum_{j=1}^{n} \hat{\lambda}_j \left( \frac{1}{n}\boldsymbol{\varepsilon}_{\hat{\theta}}^\top \hat{\phi}_j(\boldsymbol{x}) \right)^2.
$$
It should be noted that the conventional ICM test statistic also exhibits a similar discrepancy between empirical and population eigensystems, as researchers must implicitly estimate the eigenvalues and eigenfunctions of a covariance operator; see, e.g., \cite{escanciano2009lack}.

In this paper, we investigate the discrepancy between empirical and population directional components--quantified by 
$\sqrt{\hat{\lambda}_j}\,\boldsymbol{\varepsilon}_{\theta_0}^\top \hat{\phi}_j(\boldsymbol{x}) - \sqrt{\lambda_j}\,\boldsymbol{\varepsilon}_{\theta_0}^\top \phi_j(\boldsymbol{x})$--both asymptotically and non-asymptotically, and analyze its implications for the KCM test. Our analysis employs tools from the perturbation theory of compact operators.

We find that this discrepancy has a negligible impact on the asymptotic distribution of the KCM test statistic under the null hypothesis. However, under a fixed alternative, it can introduce a non-negligible bias that adversely affects the test's power. More importantly, we derive non-asymptotic results on the estimation error of the directional components. Our analysis reveals that this error is critically governed by the local eigengap $\min\{\lambda_{j-1} - \lambda_j, \lambda_j - \lambda_{j+1}\}$ for $j \geq 2$, and $\lambda_1 - \lambda_2$ for $j=1$. When the eigengap is small, even minor perturbations in the kernel matrix can induce large deviations in the estimated eigenvectors; conversely, a larger eigengap ensures greater stability in eigenvector estimation. Hence, attempting to detect deviations from the null in directions associated with small eigenvalues is unlikely to reflect genuine model misspecification; instead, it tends to introduce noise that obscures the underlying signal.

This insight motivates a simple yet effective power-boosting strategy for the KCM test: we propose truncating the empirical directional components associated with small eigenvalues, which are most vulnerable to estimation error. By discarding these noisy, low-signal directions, we aim to improve the signal-to-noise ratio of the test statistic, and thereby enhance its finite-sample power. Leveraging recent advances in the theory of smooth radial kernels \citep{belkin2018approximation}, we provide a principled theoretical justification for selecting the truncation threshold based on the decay rate of the kernel's eigenvalues. We further demonstrate, --through extensive simulation studies, --that this truncation strategy substantially improves both the test's power against a broad class of alternatives and its finite-sample size control.

Our conclusion regarding the omnibus nature of the KCM test echoes that of \cite{escanciano2009lack}, who demonstrate that the power of the ICM test is effectively concentrated in a finite-dimensional subspace of alternatives—nevertheless, the differences between our work and \cite{escanciano2009lack} are substantial. First, we arrive at this conclusion by analyzing the discrepancy between the empirical and population directional components, whereas \cite{escanciano2009lack} studies the power properties of the ICM test through the asymptotic local power function (ALPF). Second, our analysis reveals that the components associated with small eigenvalues in the test statistic are unlikely to capture genuine model misspecification; instead, they introduce noise that obscures the underlying signal. In contrast, the ALPF analysis in \cite{escanciano2009lack} establishes only that the test has limited ability to detect alternatives in these directions and does not address the noise amplification issue inherent in estimating low-eigenvalue components. Third, our analysis is mainly non-asymptotic, providing explicit conditions under which the discrepancy between empirical and population directional components can be substantial (or negligible) in finite samples. By contrast, \cite{escanciano2009lack} focuses exclusively on asymptotic power properties.

The paper is organized as follows. In Section \ref{sec:RKHS_intro}, we provide a brief overview of RKHS and related concepts needed for our analysis. Section \ref{sec:hidden_effect} presents our main theoretical results on the estimation error of the directional components, along with their implications for the KCM test. In Section \ref{sec:boosting_strategy}, we introduce our power-boosting strategy based on truncation and discuss practical guidelines for its implementation. Section \ref{sec:simulation} reports simulation results and empirical applications to illustrate the effectiveness of our proposed methods. Finally, Section \ref{sec:conclusion} concludes the paper. All proofs and additional technical details are relegated to the Appendix.

Throughout the paper, unless specified otherwise, an operator is assumed to be positive definite, self-adjoint, and compact. Additionally, we make the following assumptions.

\begin{assumption}
    \label{assumption:regular}
    The data $\{(x_i, y_i)\}_{i=1}^n$ are independent and identically distributed (i.i.d.) samples drawn from the joint distribution of $(X,Y)$. The parameter space $\Theta$ is a compact subset of $\mathbb{R}^d$, and the true parameter $\theta_0$ lies in the interior of $\Theta$. The regression function $\mathcal{M}_\theta(x)$ is twice continuously differentiable with respect to $\theta$ for each fixed $x$.
\end{assumption}

Assumption~\ref{assumption:regular} is standard in the literature on specification testing; see, for example, \cite{escanciano2024gaussian}.

\begin{assumption}
    \label{assumption:kernel}
    The kernel function $k(\cdot, \cdot)$ is a symmetric, positive-definite Mercer kernel, and the associated RKHS $\mathcal{H}_k$ is separable. In addition, the kernel is integrally strictly positive definite (ISPD), i.e.,
    $$ 
        \int_{\mathcal{X}} \int_{\mathcal{X}} k(s, t) f(s) f(t) \, dF_X(s) \, dF_X(t) > 0, \quad \forall f \in L_2(\mathbb{P}) \setminus \{0\},
    $$
    where $\mathcal{X} \subset \mathbb{R}^d$ is the support of the distribution of $X$.
\end{assumption}

Assumption~\ref{assumption:kernel} ensures that the kernel is sufficiently rich to capture all relevant features of the data distribution. The ISPD condition is crucial for the KCM test to be equivalent to the original conditional moment restriction; see \cite{muandet2020kernel} for details. Common examples of ISPD kernels include the Gaussian and Laplace kernels.

We adopt the following notation throughout the paper. For an operator $\mathcal{T}$ on a Hilbert space, we use $\|\mathcal{T}\|_{\mathrm{op}}$ and $\|\mathcal{T}\|_{\mathrm{HS}}$ to denote its operator and Hilbert--Schmidt norms, respectively. The eigenvalues of the operator $\mathcal{T}$ are denoted by $\{\lambda_j(\mathcal{T})\}_{j \geq 1}$. The tensor product of two functions (or vectors) $f$ and $g$ is denoted by $f \otimes g$, defined as $(f \otimes g)(h) = \langle g, h \rangle f$ for any $h$ in the corresponding Hilbert space. The identity operator on a Hilbert space is denoted by $\boldsymbol{I}$.
\section{Introduction to RKHS}
\label{sec:RKHS_intro}
This section provides a concise introduction to Reproducing Kernel Hilbert Spaces (RKHS), which play a central role in various areas of machine learning and statistics, particularly in kernel methods. For rigorous and comprehensive treatments of this topic, we refer readers to \cite{wendland2004scattered,rosasco2010learning,muandet2017kernel}.

Let $\mathcal{S} \subset \mathbb{R}^d$ be a non-empty set. A kernel $k: \mathcal{S} \times \mathcal{S} \to \mathbb{R}$ is a function that satisfies, for all $s, t \in \mathcal{S}$,
$$ 
k(s, t) = \langle \psi(s), \psi(t) \rangle_{\mathcal{H}_k},
$$  
where $\psi: \mathcal{S} \to \mathcal{H}_k$ is a feature map, and $\mathcal{H}_k$ is a Hilbert space equipped with the inner product $\langle \cdot, \cdot \rangle_{\mathcal{H}_k}$. The feature map $\phi$ implicitly maps the input space $\mathcal{S}$ into a high-dimensional (possibly infinite-dimensional) feature space $\mathcal{H}_k$. 

A typical example of a kernel is the Gaussian kernel, defined as
$$ 
k(s, t) = \exp\left(-\gamma\|s - t\|^2\right), \quad \gamma > 0.
$$
For the special case where $\gamma = 1$ and $d = 1$, the corresponding feature map is given by
$$
\psi(t) = \left(\exp(-t^2/2), t\exp(-t^2/2), \frac{t^2\exp(-t^2/2)}{\sqrt{2}}, \ldots, \frac{t^i\exp(-t^2/2)}{\sqrt{i!}}, \ldots\right),
$$
which maps $t$ to an infinite-dimensional space. Kernels satisfying this inner product condition are called positive definite, meaning that for any $c_i \in \mathbb{R}$, $t_i \in \mathcal{S}$, and $i = 1, \ldots, n$, the following inequality holds:
$$ 
\sum_{i=1}^n \sum_{j=1}^n c_i c_j k(t_i, t_j) \geq 0.
$$

The Moore--Aronszajn theorem guarantees that for every positive definite kernel $k$, there exists a unique RKHS $\mathcal{H}_k$ associated with it. This RKHS, $\mathcal{H}_k$, is a Hilbert space consisting of functions $f: \mathcal{S} \to \mathbb{R}$ such that, for all $t \in \mathcal{S}$, the (Dirac) evaluation functional $E_t: \mathcal{H}_k \to \mathbb{R}$, defined by $E_t(f) = f(t)$, is continuous. This continuity ensures the existence of a reproducing kernel, which satisfies the reproducing property:
$$ 
f(t) = \langle f, k(\cdot, t) \rangle_{\mathcal{H}_k}, \quad \forall f \in \mathcal{H}_k, \; t \in \mathcal{S}.
$$
This property allows the evaluation of functions in the RKHS using the kernel function, simplifying various computations. In particular, the kernel function itself satisfies
$$ 
k(s, t) = \langle k(\cdot, s), k(\cdot, t) \rangle_{\mathcal{H}_k}, \quad \forall s, t \in \mathcal{S}.
$$

Given a probability measure $\mathbb{P}$ on $\mathcal{S}$, we can define the integral operator $L_k: L_2(\mathbb{P}) \to L_2(\mathbb{P})$ associated with the kernel $k$ as  
$$ 
L_k f(\cdot) = \int k(\cdot, t) f(t) d\mathbb{P}(t), \quad f \in L_2(\mathbb{P}).
$$
It is straightforward to verify that $L_k$ is self-adjoint and compact when the kernel is continuous. Note that while $f \in L_2(\mathbb{P})$ needs to be defined only on the support of $\mathbb{P}$, $L_k f$ is well-defined on $\mathcal{S}$. In many cases, the difference between the support of $\mathbb{P}$ and $\mathcal{S}$ is not negligible. Moreover, it can be shown that $L_k f \in \mathcal{H}_k$ for any $f \in L_2(\mathbb{P})$, and a function $f \in \mathcal{H}_k$ gives rise to a function in $L_2(\mathbb{P})$ by restricting it to the support of $\mathbb{P}$. We define the restriction operator $\mathcal{R}: \mathcal{H}_k \to L_2(\mathbb{P})$ such that $\mathcal{R}f(x) = f(x), \forall x \in \text{supp}(\mathbb{P}) = \mathcal{X}$. Furthermore, it can be shown by an extension of the argument in \cite{wendland2004scattered} (Proposition 10.28) that $\mathcal{R}$ is the adjoint of the kernel integral operator $R^* = L_k:L_2(\mathbb{P}) \to \mathcal{H}_k$. Specifically, for any $f \in \mathcal{H}_k$ and $g \in L_2(\mathbb{P})$, we have
$$ 
    \langle \mathcal{R} f, g \rangle_{L_2(\mathbb{P})} = \langle f, L_k g \rangle_{\mathcal{H}_k}.
$$
Moreover, it turns out that the square root $(R^*)^{1/2}=L_k^{1/2}$ exists and is isometric embedding of $L_2(\mathbb{P})$ into $\mathcal{H}_k$. In fact, we have 
$$ 
    \langle L_k^{1/2} f, L_k^{1/2} g \rangle_{\mathcal{H}_k} = \langle f, g \rangle_{L_2(\mathbb{P})}, \quad \forall f, g \in L_2(\mathbb{P}),
$$ 
In addition, we have
$$ 
    \mathcal{R} \mathcal{R}^* = L_k:L_2(\mathbb{P})\to L_2(\mathbb{P}), \quad \text{and} \quad \mathcal{R}^* \mathcal{R} = \Sigma = \mathbb{E}\left(k(X,\cdot) \otimes k(X,\cdot) \right): \mathcal{H}_k \to \mathcal{H}_k,
$$ 
with $\Sigma$ being the (uncentered) covariance operator on $\mathcal{H}_k$. 

Let $\phi_i$ and $e_i$ be the eigenfunctions of $L_k$ and $\Sigma$, respectively, associated with eigenvalues $\lambda(L_k)_i$ and $\lambda(\Sigma)_i$. For any $i$, we have the following relationships (Proposition 8 in \cite{rosasco2010learning}):
\begin{align*}
    & \lambda(L_k)_i = \lambda(\Sigma)_i = \lambda_i, \\
    & \sqrt{\lambda_i} \phi_i(x) = e_i(x), \quad \text{ for } \mathbb{P}\text{--almost all } x \in \mathcal{X}, \\
\end{align*}   
In this paper, we assume that the eigenvalues of $L_k$ (or equivalently, $\Sigma$) are \textit{simple}, meaning $\lambda_i \neq \lambda_j$ for all $i \neq j$. The eigenvalues are ordered in descending sequence: $\lambda_1 > \lambda_2 > \cdots > 0$.

Mercer's theorem states that if the kernel $k$ is continuous, symmetric, and positive definite, then the following decomposition holds:
$$ 
    k(s, t) = \sum_{i=1}^\infty \lambda_i \phi_i(s) \phi_i(t), \quad \forall s, t \in \mathcal{S},
$$
where the series converges absolutely and uniformly. The eigenfunctions $\{\phi_i\}_{i=1}^\infty$ form an orthonormal basis of $L_2(\mathbb{P})$. A useful formula derived from this decomposition is the \textit{Nystr\"om extension}, which expresses the eigenfunctions outside the support of $\mathbb{P}$ as:
$$ 
    \phi_i(z) = \frac{1}{\lambda_i} \int k(z, t) \phi_i(t) d\mathbb{P}(t), \quad \forall z \in \mathcal{S}.
$$ 

It is important to note that the eigenfunctions $\{\phi_i\}_{i=1}^\infty$ are technically elements of $L_2(\mathbb{P})$. However, since the image of $L_k$ lies in $\mathcal{H}_k$, we can also view $\phi_i$ as elements of $\mathcal{H}_k$. Thus, each eigenfunction $\phi_i$ corresponds to two representations: one in $L_2(\mathbb{P})$, denoted by $\phi_{i,L_2(\mathbb{P})}$, and the other in $\mathcal{H}_k$, denoted by $\phi_{i,\mathcal{H}_k}$. These representations coincide on the support of $\mathbb{P}$, i.e., $\phi_{i,L_2(\mathbb{P})}(x) = \phi_{i,\mathcal{H}_k}(x)$ for all $x \in \mathcal{X}$, but they have different norms in their respective spaces. Specifically (see, e.g., \cite{rosasco2010learning} for details), the following relationships hold:
$$ 
    \mathcal{R}\phi_{i,\mathcal{H}_k} = \phi_{i,L_2(\mathbb{P})}, \quad \lambda_i^{-1}L_k \phi_{i,L_2(\mathbb{P})} = \phi_{i,\mathcal{H}_k}.
$$ 
For simplicity, we use the same notation $\phi_i$ for both representations, while keeping their respective norms in note.

Given independent and identically distributed (i.i.d.) samples $\{x_\alpha\}_{\alpha=1}^n$ drawn from $\mathbb{P}$, we can define the empirical covariance operator $\Sigma_n: \mathcal{H}_k \to \mathcal{H}_k$ as
$$ 
    \Sigma_n = \frac{1}{n}\sum_{\alpha=1}^n k(x_\alpha, \cdot) \otimes k(x_\alpha, \cdot).
$$

Consequently, define the empirical restriction operator $\mathcal{R}_n: \mathcal{H}_k \to \mathbb{R}^n$ as
$$ 
    \mathcal{R}_n f = (\sqrt{n})^{-1}(f(x_1), \ldots, f(x_n))^\top, \quad f \in \mathcal{H}_k,
$$
which leads to 
$$ 
    \mathcal{R}_n \mathcal{R}_n^* = \mathbf{K}/n, \quad \text{and} \quad \mathcal{R}_n^* \mathcal{R}_n = \Sigma_n,
$$
where $\mathcal{R}_n^*$ is the adjoint operator of $\mathcal{R}_n$, $\mathbf{K} \in \mathbb{R}^{n \times n}$, with the entries given by $\mathbf{K}_{\alpha\beta} = k(x_\alpha, x_\beta)$. The scaled kernel matrix $\mathbf{K}/n$ can be interpreted as a discrete approximation of the integral operator $L_k$, where the probability measure $\mathbb{P}$ is replaced by the empirical distribution $\mathbb{P}_n = 1/n\sum_{\alpha=1}^n \delta_{x_\alpha}$, with $\delta_{x}$ denoting the Dirac measure at $x$.

\section{Estimation Effect from Eigensystem}
\label{sec:hidden_effect}
Throughout this section, we assume that the true parameter $\theta_0$ is known. This assumption allows us to isolate and examine the impact of estimating the integral operator. We first investigate the origin and structures of the estimation effect in KCM-type statistics, followed by an analysis of its asymptotic behavior. Finally, we discuss the implications of these findings for finite-sample performance.

Recall the quantity of interest in its population form is:
\begin{align*}
    T  = \mathbb{E}\left[\varepsilon_{\theta_0} k(X,X') \varepsilon_{\theta_0}'\right] = \sum_{j=1}^{\infty} \lambda_j \mathbb{E}\left[\varepsilon_{\theta_0} \phi_j(X) \phi_j(X') \varepsilon_{\theta_0}'\right],
\end{align*}
where the second equality follows from Mercer's decomposition of the kernel $k$. The empirical counterpart is given by
\begin{align*}
    T_n  = \frac{1}{n^2} \sum_{\alpha,\beta=1}^n \varepsilon_{\theta_0,x_\alpha} k(x_\alpha, x_\beta) \varepsilon_{\theta_0,x_\beta}  = \frac{1}{n} \sum_{j=1}^{n} \hat{\lambda}_j \boldsymbol{\varepsilon}_{\theta_0}^\top \boldsymbol{U}_j \boldsymbol{U}_j^\top \boldsymbol{\varepsilon}_{\theta_0},
\end{align*}
where $\{\hat{\lambda}_j, \boldsymbol{U}_j\}_{j=1}^n$ are the eigenvalue-eigenvector pairs of the scaled kernel matrix $\mathbf{K}/n$, and $\boldsymbol{\varepsilon}_{\theta_0} = (\varepsilon_{\theta_0,1}, \ldots, \varepsilon_{\theta_0,n})^\top$. The estimation effects arise from replacing the population unknown quantities $\lambda_j$ and $\phi_j$ with their empirical counterparts $\hat{\lambda}_j$ and $\boldsymbol{U}_j$, which are stochastic and depend on the sample $\{x_\alpha\}_{\alpha=1}^n$.

To analyze the estimation effect, we begin with a singular value decomposition of the restriction operator $\mathcal{R}_n:\mathcal{H}_k \to \mathbb{R}^n$ (Proposition 9 in \cite{rosasco2010learning}):
$$ 
    \mathcal{R}_n = \sum_{j=1}^n \sqrt{\hat{\lambda}_j} \boldsymbol{U}_j \otimes \hat{e}_j,
$$
with $\langle \hat{e}_i, \hat{e}_j \rangle_{\mathcal{H}_k} = \delta_{ij}$, $ 
    \hat{e}_j = \hat{\lambda}_j^{-1/2} \mathcal{R}_n^* \boldsymbol{U}_j,
$ and $ 
    \boldsymbol{U}_j = \hat{\lambda}_j^{-1/2} \mathcal{R}_n \hat{e}_j.
$

The Nystr\"om approximation of the eigenfunction $\phi_j$ is given by 
$$ 
    \hat{\phi}_j(\cdot) = \frac{1}{\sqrt{\hat{\lambda}_j}} \hat{e}_j(\cdot) =\frac{1}{\hat{\lambda}_j} \mathcal{R}_n^* \boldsymbol{U}_j,
$$ 
and 
\begin{align*}
    \hat{\phi}_j(\boldsymbol{x})& =\frac{\sqrt{n}}{\sqrt{n}}(\hat{\phi}_j(x_1), \ldots, \hat{\phi}_j(x_n))^\top \\
    & = \sqrt{n} \mathcal{R}_n \hat{\phi}_j \\
    & = \frac{\sqrt{n}}{\hat{\lambda}_j} \mathcal{R}_n \mathcal{R}_n^* \boldsymbol{U}_j \\
    & = \frac{\sqrt{n}}{\hat{\lambda}_j} \frac{\boldsymbol{K}}{n} \boldsymbol{U}_j = \sqrt{n}\boldsymbol{U}_j.
\end{align*}

The (scaled) KCM-type statistics commonly found in the literature can be expressed as:
\begin{align*}
    n T_n &= \sum_{j=1}^n  \left[\frac{1}{\sqrt{n}} \boldsymbol{\varepsilon}_{\theta_0}^\top \sqrt{\hat{\lambda}_j}\hat{\phi}_j(\boldsymbol{x})\right]^2\\ 
    & = \sum_{j=1}^n  \left[\frac{1}{\sqrt{n}} \boldsymbol{\varepsilon}_{\theta_0}^\top \hat{e}_j(\boldsymbol{x})\right]^2\\
    & = \sum_{j=1}^n \left[\frac{1}{\sqrt{n}} \boldsymbol{\varepsilon}_{\theta_0}^\top e_j(\boldsymbol{x}) + \frac{1}{\sqrt{n}} \boldsymbol{\varepsilon}_{\theta_0}^\top (\hat{e}_j - e_j)(\boldsymbol{x})\right]^2.
\end{align*}
We aim to analyze the impact of replacing the true quantities ($\{e_j\}_{j=1}^n$) with their empirical counterparts ($\{\hat{e}_j\}_{j=1}^n$) in both asymptotic and finite-sample settings. The primary tool for this analysis is the analytic perturbation theory for linear operators. For a comprehensive reference, see \cite{kato2013perturbation}, and for a more accessible introduction, refer to \cite{hsing2015theoretical}.

The following theorem will be used throughout this section:
\begin{theorem}
    \label{thm:upper_bound_eigenfunction}
    Let $\mathcal{T}$ and its approximation $\mathcal{\hat{T}}$ be positive definite, compact and self-adjoint operators on a Hilbert space $\mathcal{H}$, with eigenvalues $\{\lambda_j\}_{j=1}^\infty$ and $\{\hat{\lambda}_j\}_{j=1}^\infty$, respectively. The corresponding eigenfunctions are $\{e_j\}_{j=1}^\infty$ and $\{\hat{e}_j\}_{j=1}^\infty$. Define $\Delta = \mathcal{\hat{T}} - \mathcal{T}$. If $\langle e_j, \hat{e}_j \rangle_{\mathcal{H}} \geq 0$, then:
    \begin{equation}
        \label{eq:general_eigenfunction_perturbation}
        \hat{e}_j - e_j = \sum_{i \neq j} (\hat{\lambda}_j - \lambda_i)^{-1} \mathcal{P}_i \Delta \hat{e}_j + \mathcal{P}_j (\hat{e}_j - e_j),
    \end{equation}
    where $\mathcal{P}_j$ is the projection operator associated with the eigenfunction $e_j$:
    $$ 
        \mathcal{P}_j = e_j \otimes e_j.
    $$  
    Let $\eta_j = \frac{1}{2} \inf_{i \neq j} |\lambda_j - \lambda_i|$ and $\delta_j = \| \Delta \|_{op} / \eta_j$. If $\delta_j < 1$, then:
    $$ 
        \| e_j - \hat{e}_j - \mathcal{D}_j \Delta e_j \|_{\mathcal{H}} \leq \Psi(\delta_j) \delta_j^2,
    $$  
    for some finite function $\Psi$, where $\Psi(\delta) \to 1^+$ as $\delta \to 0^+$. Here, the term:
    $$ 
        \mathcal{D}_j = \sum_{i \neq j} \frac{1}{\lambda_i - \lambda_j} \mathcal{P}_i
    $$
    is called the reduced resolvent of $\mathcal{T}$. 
\end{theorem}
\begin{proof}
    This is a simplified version of Theorem 5.1.8 in \cite{hsing2015theoretical}.
\end{proof}
\begin{remark}
    In our application, we set $\mathcal{T} = \Sigma$ and $\mathcal{\hat{T}} = \hat{\Sigma}$.
\end{remark}

\begin{remark}
    When the eigenvalues of $\mathcal{T}$ are simple, i.e., $\lambda_1 > \lambda_2 > \ldots > \lambda_n > \ldots > 0$, we define $\eta_j = \frac{1}{2} \min\{\lambda_{j-1} - \lambda_j, \lambda_j - \lambda_{j+1}\}$ for $j > 1$, and $\eta_1 = \frac{1}{2} (\lambda_1 - \lambda_2)$. Notably, $\eta_j > 0$ for all $j$. Since $\| \Delta \|_{op}=\| \hat{\Sigma}-\Sigma \|_{op} = O_p(n^{-1/2})$ (see Lemma \ref{lemma:HS_bound} in the Appendix), it follows that $\delta_j = O_p(n^{-1/2})$. Consequently, the condition $\delta_j < 1$ is automatically satisfied as $n \to \infty$.

    In analytic perturbation theory, the term $\mathcal{D}_j \Delta e_j$ is commonly referred to as the first-order approximation of the eigenfunction perturbation, see, e.g., \cite{griffiths2018introduction}. Provided $\delta_j <1$, the eigenfunction $e_j$ can be approximated by:
    \begin{equation}
        \label{eq:first_order_approximation}
        e_j \approx \hat{e}_j + \mathcal{D}_j \Delta e_j = \hat{e}_j + \sum_{i \neq j} \frac{\langle e_i, \Delta e_j \rangle_{\mathcal{H}}}{\lambda_i - \lambda_j} e_i.
    \end{equation}
\end{remark}

\subsection{Asymptotic Analysis of Estimation Effect}
The following lemma is required for analyzing the asymptotic effects of the estimation. 
\begin{lemma}
\label{lemma:analytic_perturbation} 
    Let the kernel $k$ be continuous, symmetric, and positive definite, and the eigenvalues of the associated integral operator $L_k$ be simple. Then,
    $$ 
        (\hat{e}_j - e_j)(x) = \sqrt{\lambda_j} \sum_{i \neq j} \frac{\lambda_i}{\lambda_j-\lambda_i} \left(\frac{1}{n}\sum_{\alpha=1}^n \phi_i(x_\alpha)\phi_j(x_\alpha)\right) \phi_i(x) + O_p(n^{-1}), \, \forall x \in \mathbb{R}^d.
    $$
\end{lemma}
The next theorem is a direct application of Lemma~\ref{lemma:analytic_perturbation}. 
\begin{theorem}
    \label{thm:residual_asymptotics}
    Under the same conditions as in Lemma~\ref{lemma:analytic_perturbation}, we have
    $$ \frac{1}{\sqrt{n}} \boldsymbol{\varepsilon}_{\theta_0}^\top (\hat{e}_j- e_j)(\boldsymbol{x}) = \begin{cases}
            o_p(1), & \text{if null hypothesis holds}, \\
           
         O_p(1), & \text{if alternative hypothesis holds}.
        \end{cases}.
    $$
    In particular, under the alternative hypothesis,
    $$ 
        \frac{1}{\sqrt{n}} \boldsymbol{\varepsilon}_{\theta_0}^\top (\hat{e}_j- e_j)(\boldsymbol{x}) \overset{d}{\longrightarrow} \sqrt{\lambda_j} \sum_{i \neq j} \frac{\lambda_i}{\lambda_j-\lambda_i} \mathbb{E}[\varepsilon_{\theta_0} \phi_i(X)] W_{ij},
    $$
    where $W_{ij} \sim \mathcal{N}(0,\sigma_{ij}^2)$ with $\sigma_{ij}^2 = \mathrm{Var}(\phi_i(X)\phi_j(X))$.
\end{theorem}
The above theorem indicates that the estimation effect is asymptotically negligible under the null hypothesis:
\begin{align*}
    n T_n &= \sum_{j=1}^n \left[\frac{1}{\sqrt{n}} \boldsymbol{\varepsilon}_{\theta_0}^\top e_j(\boldsymbol{x})\right]^2 + o_p(1)\\
    & = \sum_{j=1}^n \lambda_j \left[\frac{1}{\sqrt{n}} \boldsymbol{\varepsilon}_{\theta_0}^\top \phi_j(\boldsymbol{x})\right]^2 + o_p(1)
\end{align*}
The following null asymptotic distribution result is a direct consequence of the above equation. 
\begin{corollary}
    \label{cor:null_distribution}
    Under the null hypothesis and the same conditions as Lemma~\ref{lemma:analytic_perturbation}, as well as the kernel being ISPD, we have 
    $$ 
        n T_n \overset{d}{\longrightarrow} \sum_{j=1}^\infty \lambda_j W_j^2,
    $$ 
    where $W_j \sim \mathcal{N}(0,\mathrm{Var}(\varepsilon_{\theta_0}\phi_j(X)))$ and have covariance structure $\mathrm{Cov}(W_i, W_j) = \mathbb{E}(\varepsilon_{\theta_0}^2 \phi_i(X)\phi_j(X))$ for $i \neq j$.
\end{corollary}

However, the estimation effect is not negligible and can be considered as noise that may obscure the true signal. Specifically, $(n)^{-1/2} \boldsymbol{\varepsilon}_{\theta_0}^\top (\hat{e}_j-e_j)(\boldsymbol{x})$ is distributed around zero:
$$ 
    \frac{1}{n} \boldsymbol{\varepsilon}_{\theta_0}^\top (\hat{e}_j-e_j)(\boldsymbol{x}) \overset{p}{\longrightarrow} 0.
$$
Despite the presence of this noise, it is straightforward to observe that $nT_n$ still diverges to infinity under the fixed alternative hypothesis, ensuring the consistency of the test statistic.

\subsection{Finite-Sample Analysis of Estimation Effect}

For a finite and fixed sample size $n$, the sufficient condition $\delta_j = \| \hat{\Sigma} - \Sigma \|_{op} / \eta_j < 1$ becomes crucial for the validity of the first-order approximation of eigenfunction perturbation. When this condition holds, the estimation effect is well-controlled, in the sense that small changes in the data result in only minor deviations in the eigenfunction estimate. Conversely, if this condition is violated, even small changes in the data can cause the eigenfunction estimate to deviate significantly from the true eigenfunction. This issue is examined from two complementary perspectives: numerical and geometric.

First, note that Equation (\ref{eq:general_eigenfunction_perturbation}) holds regardless of the value of $\delta_j$. When $\delta_j < 1$, it can be shown that the interval $\mathcal{I}_j = (\lambda_j - \eta_j, \lambda_j + \eta_j)$ contains only $\lambda_j$ and $\hat{\lambda}_j$, and no other eigenvalues of $\Sigma$ or $\hat{\Sigma}$\footnote{It is straightforward to verify that $\lambda_j$ lies within this interval, while no other eigenvalues of $\Sigma$ are included. To confirm that $\hat{\lambda}_j$ is also within the interval, one can refer to Lemma \ref{lemma:upper_bound_operator} in the Appendix. Finally, a proof by contradiction shows that no other eigenvalues of $\hat{\Sigma}$ are included. Assume, for some $i$, that $\hat{\lambda}_i \in \mathcal{I}_j$. Then $|\lambda_j - \hat{\lambda}_i| < \eta_j$. However, by Lemma \ref{lemma:upper_bound_operator}, we know $|\lambda_i - \hat{\lambda}_i| \leq \|\hat{\Sigma} - \Sigma\|_{op} < \eta_j$. Using the triangle inequality, we obtain $|\lambda_j - \lambda_i| \leq |\lambda_j - \hat{\lambda}_i| + |\hat{\lambda}_i - \lambda_i| < \eta_j + \eta_j = 2\eta_j$, which contradicts the definition of $\eta_j$.}. In this case, Equation (\ref{eq:first_order_approximation}) is guaranteed to hold, ensuring that small changes in the data result in only minor changes in the eigenfunction estimate:
$$ 
    \mathcal{D}_j (\hat{\Sigma} - \Sigma) e_j = \sqrt{\lambda_j} \sum_{i \neq j} \frac{\lambda_i}{\lambda_j - \lambda_i} \left(\frac{1}{n} \sum_{\alpha=1}^n \phi_i(x_\alpha) \phi_j(x_\alpha)\right) \phi_i \approx \boldsymbol{0} \in \mathcal{H}_k,
$$
as $n^{-1} \sum_{\alpha=1}^n \phi_i(x_\alpha) \phi_j(x_\alpha) \approx 0$ for any $i \neq j$. Furthermore, Theorem~\ref{thm:upper_bound_eigenfunction} provides the bound:
$$ 
    \| e_j - \hat{e}_j - \mathcal{D}_j (\hat{\Sigma} - \Sigma) e_j \|_{\mathcal{H}_k} \leq \Psi(\delta_j) \delta_j^2.
$$

On the other hand, if $\delta_j \geq 1$, the estimated eigenvalue $\hat{\lambda}_j$ may fall outside the interval $\mathcal{I}_j$ and instead lie within a neighboring interval $\mathcal{I}_i = (\lambda_i - \eta_i, \lambda_i + \eta_i)$ for some $i \neq j$. In this scenario, the term $(\hat{\lambda}_j - \lambda_i)^{-1}$ in Equation \ref{eq:general_eigenfunction_perturbation} can become large or even undefined (if $\hat{\lambda}_j = \lambda_i$). Consequently, the eigenfunction estimate $\hat{e}_j$ may deviate significantly from the true eigenfunction $e_j$.

The geometric argument involves the contour integral representation of the projection operator $\mathcal{P}_j = e_j \otimes e_j$. To begin, recall that for a bounded and self-adjoint operator such as $\Sigma$, the spectral theorem states: 
$$ 
    \Sigma = \sum_{j=1}^\infty \lambda_j \mathcal{P}_j.
$$
Remarkably, the projection operator $\mathcal{P}_j$ can be expressed as a contour integral:
$$ 
    \mathcal{P}_j = -\frac{1}{2\pi \ii} \oint_{\Gamma_j} R(z) \, dz,
$$
where $R(z) = (\Sigma - zI)^{-1} = \sum_{j=1}^\infty (\lambda_j - z)^{-1} \mathcal{P}_j$ is the resolvent operator of $\Sigma$, and $\Gamma_j \subset \mathbb{C}$ is a closed contour in the complex plane that encloses only the eigenvalue $\lambda_j$, excluding all other eigenvalues of $\Sigma$. A common choice for $\Gamma_j$ is a circle centered at $\lambda_j$ with radius $\eta_j$. While the integral of an operator-valued function may initially seem unfamiliar, it is defined analogously to the integral of a real-valued function. Despite its technical appearance, the above equation enables a much simpler analysis of eigenprojections compared to other seemingly more direct methods.

The condition $\delta_j < 1$ ensures that the contour $\Gamma_j$ also encloses only the estimated eigenvalue $\hat{\lambda}_j$ of $\hat{\Sigma}$, excluding all other eigenvalues of either $\Sigma$ or $\hat{\Sigma}$. Consequently, the difference between the true projection operator and the estimated projection operator $\hat{\mathcal{P}}_j = \hat{e}_j \otimes \hat{e}_j$ can be expressed as:
$$ 
    \hat{\mathcal{P}}_j - \mathcal{P}_j = -\frac{1}{2\pi \ii} \oint_{\Gamma_j} (\hat{R}(z) - R(z)) \, dz,
$$
where $\hat{R}(z) = (\hat{\Sigma} - zI)^{-1}$ is the resolvent operator of $\hat{\Sigma}$. 

\begin{figure}[htbp]
    \centering
        \includegraphics[width=0.8\textwidth]{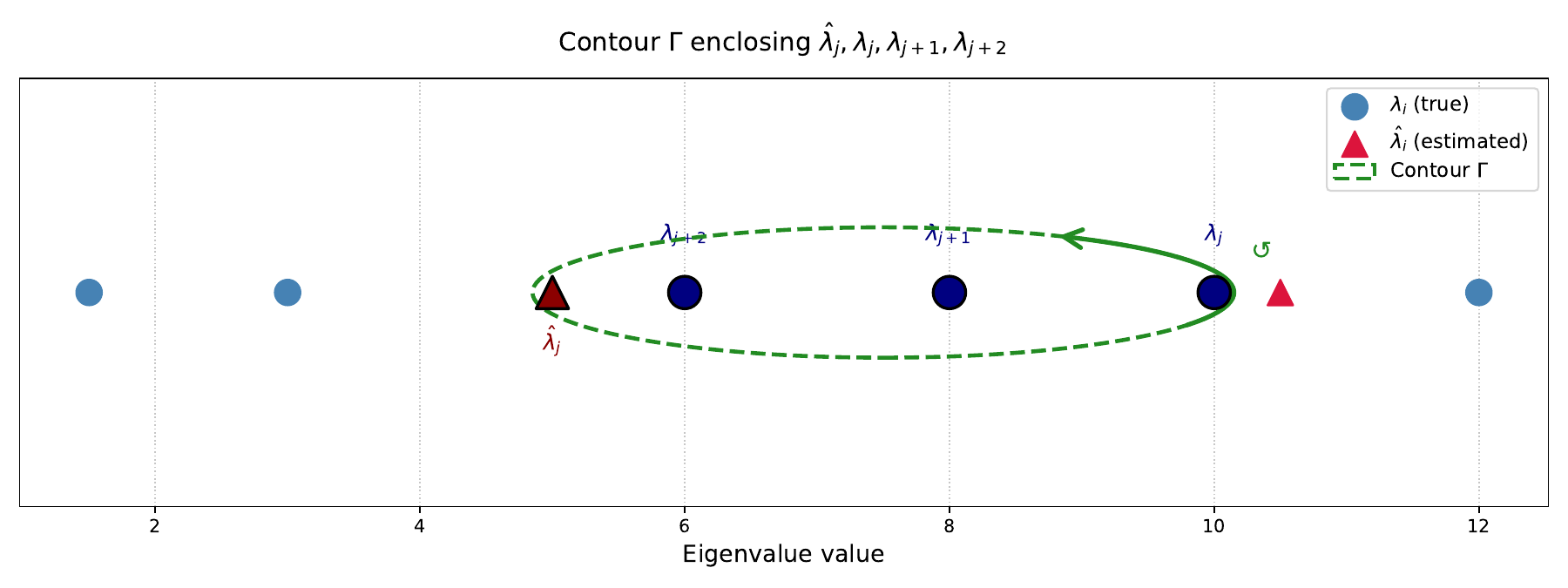}
    \caption{Eigenvalues and the enclosing contour $\Gamma$.}
    \label{fig:contour_plot}
\end{figure}

When $\delta_j \geq 1$, the smallest contour $\Gamma$ that encloses both $\lambda_j$ and $\hat{\lambda}_j$ may also enclose other eigenvalues of either $\Sigma$ or $\hat{\Sigma}$. For instance, it could happen that $\hat{\lambda}_j, \lambda_j, \lambda_{j+1}, \lambda_{j+2} \in \Gamma$, see Figure \ref{fig:contour_plot}. 
In this case, the contour integral representation of the difference becomes:
$$ 
    -\frac{1}{2\pi \ii} \oint_{\Gamma} (\hat{R}(z) - R(z)) \, dz = \hat{\mathcal{P}}_j - \left(\mathcal{P}_{j} + \mathcal{P}_{j+1} + \mathcal{P}_{j+2}\right),
$$
where the integral of $\hat{R}(z)$ yields $\hat{\mathcal{P}}_j$ since $\Gamma$ only encloses $\hat{\lambda}_j$, while the integral of $R(z)$ yields the sum of the projection operators corresponding to all eigenvalues enclosed by $\Gamma$. As a result, the eigenfunction estimate $\hat{e}_j$, which lies in the range of $\hat{\mathcal{P}}_j$, may be close to a linear combination of $e_{j}, e_{j+1}$, and $e_{j+2}$, rather than being close to $e_j$ alone.

Geometrically, this implies that when $\delta_j < 1$, the quantity $(n)^{-1/2} \boldsymbol{\varepsilon}_{\theta_0}^\top \hat{e}_j (\boldsymbol{x}) = (n)^{-1/2} \boldsymbol{\varepsilon}_{\theta_0}^\top e_j (\boldsymbol{x}) + (n)^{-1/2} \boldsymbol{\varepsilon}_{\theta_0}^\top (\hat{e}_j - e_j)(\boldsymbol{x})$ effectively captures the true signal $(n)^{-1/2} \boldsymbol{\varepsilon}_{\theta_0}^\top e_j (\boldsymbol{x})$ in a direction aligned with $e_j$, with only minor interference from the estimation effect. In contrast, when $\delta_j \geq 1$, the noise term $(n)^{-1/2} \boldsymbol{\varepsilon}_{\theta_0}^\top (\hat{e}_j - e_j)(\boldsymbol{x})$ can become substantial, potentially overwhelming and obscuring the true signal in the direction of $e_j$.

\section{Truncating the Test Statistic}
\label{sec:boosting_strategy}
Our previous analysis suggests that an effective test statistic should assign greater weight to directions with $\delta_j < 1$. For a fixed sample size $n$, this is essentially equivalent to prioritizing directions with larger $\eta_j$. Building on this insight, we propose enhancing the finite-sample performance of KCM-type tests by simply truncating the directions.

When the eigenvalues of the operator $\Sigma$ (or $L_k$) are simple and sorted in descending order, the leading eigenvalues typically correspond to larger $\eta_j$ values. Thus, a natural approach to improve the finite-sample performance of the KCM test is to truncate the directions, including only the first $J(n)$ directions in the test statistic, where $J(n)$ increases with the sample size $n$. Specifically, we consider the following truncated KCM test statistic:
\begin{equation}
    \label{eq:truncated_KCM}
    n T_{n,J(n)} = \sum_{j=1}^{J(n)} \left( \frac{1}{\sqrt{n}} \boldsymbol{\varepsilon}_{\theta_0}^\top \hat{e}_j (\boldsymbol{x}) \right)^2,
\end{equation}  
where $\hat{e}_j(\boldsymbol{x})$ is computed using the kernel matrix as previously described. The choice of $J(n)$ is critical and involves balancing the test's omnibus property against noise control. Some words on the null asymptotic distribution of the truncated KCM statistic $n T_{n,J(n)}$ are in order. Since $J(n)$ diverges to infinity as $n \to \infty$, the asymptotic distribution of $n T_{n,J(n)}$ under the null hypothesis is the same as that of the full KCM statistic $n T_n$. Standard bootstrap procedures can also be applied to approximate the null distribution of $n T_{n,J(n)}$. 

Selecting the truncation level $J(n)$ is critical to balancing the test statistic's ability to detect a broad range of alternatives (omnibus property) against its robustness to noise. While an optimal, data-driven approach to determine $J(n)$ would be ideal, this remains an open problem for future investigation. Below, we outline key findings that could inform the development of effective truncation strategies in future research.

\begin{assumption}
\label{assumption:kernel_decay}
Let $k(s,t)$ be a smooth radial kernel defined on $\mathbb{R}^d$ with $d$ fixed. Specifically, let $k(s,t) = h(\|s-t\|)$ and set $g(\cdot) = h(\sqrt{\cdot})$. Assume that $|g^{(l)}(r)| \leq l! M^l$ for all $l$ and $r>0$, where $g^{(l)}$ denotes the $l$-th order derivative of $g$, and $M$ is a positive constant.
\end{assumption}

Two important kernels satisfying this condition are the Gaussian kernel and the inverse multiquadratic kernel. The following lemma characterizes the eigenvalue decay rate of the integral operator $L_k$ (and $\Sigma$) associated with such kernels.

\begin{lemma}
\label{lemma:eigen_decay}
Let $\kappa = \sup_{x \in \mathcal{X} \subset \mathbb{R}^d}k(x,x)$, and let $k$ be a kernel satisfying Assumption \ref{assumption:kernel_decay}. Then for some $C,C'>0$, we have 
$$ 
    \lambda_j \leq \sqrt{\kappa} C \exp(-C' j^{1/d}), \quad \text{for all } j \geq 1.
$$ 
\end{lemma}
\begin{proof}
    See Theorem 5 in \cite{belkin2018approximation}.
\end{proof}
\begin{remark}
    Notice that all quantities in Lemma \ref{lemma:eigen_decay} are independent of the probability measure $\mathbb{P}$. In particular, when $\mathbb{P}$ is a finite measure, the integral operator $L_k$ can be viewed as a matrix. Hence, this result provides a uniform bound on the eigenvalue decay rate, independent of the size of the matrix. 
\end{remark}

It is straightforward to see that $\eta_j = (1/2) (\lambda_j - \lambda_{j+1})$ since $\exp(-C' j ^{1/d})$ is convex and decreasing in $j$. The sufficient condition $\delta_j < 1$ can be implied by\footnote{One may use the approximation $\eta_j \approx (1/2) | (d / d j) \lambda_j|>O_p(n^{-1/2})$.}  
$$ 
    C' j^{1/d} + \frac{d-1}{d} \log j < \frac{1}{2} \log n - \log d +M_0,
$$
where $M_0$ is a constant depending on $C, C'$ and $\kappa$. Solving for $j$ in terms of $n$ involves a transcendental equation, which is not straightforward. However, some intuition can be gained by considering the case $d=1$, where the condition reduces to
$$ 
    j < \frac{1}{2C'} \log n +M_0.
$$ 
Thus, even in the simple case $d=1$, the sample size $n$ required to include more directions grows exponentially, which is demanding in practice. In simulation studies, we find that setting $J(n) = 1$ with sample sizes $n=200$ and $n=400$ often yields satisfactory results.

\begin{remark}
    One might also observe the ``curse of dimensionality'' through the lens of eigenvalue decay. As $d$ increases, the eigenvalues $\lambda_j$ tend to decay more slowly, which results in a smaller spectral gap $\eta_j$, making it more challenging to satisfy the condition $\delta_j < 1$ even for small $j$.
\end{remark}

\section{Simulation Studies and Empirical Applications}
\label{sec:simulation}
In this section, we conduct extensive simulation studies to evaluate the finite-sample performance of the proposed truncating power-boosting strategy for the KCM test. We compare our methods with existing approaches in the literature across various scenarios, including different sample sizes, dimensions, and types of alternatives. Additionally, we apply our methods to real-world datasets to demonstrate their practical utility.

In practice, the residual terms $\{\varepsilon_{\theta_0,\alpha}\}_{\alpha=1}^n$ are unknown and need to be estimated by $\{\varepsilon_{\hat{\theta},\alpha}\}_{\alpha=1}^n$ using a consistent estimator $\hat{\theta}$ of $\theta_0$. Several methods exist to handle this estimation effect. A common approach is to use a Neyman orthogonal operator $\boldsymbol{\Pi}: \mathcal{H}_k \to L_2(\mathbb{P})$  on the feature map $k(x,\cdot)$ such that 
$$
\nabla_{\theta = \theta_0}\mathbb{E}[\varepsilon_{\theta} \boldsymbol{\Pi} k(x,X)]=\boldsymbol{0} \in \mathbb{R}^d;
$$ 
see, e.g., \cite{escanciano2014specification,sant2019specification,escanciano2024gaussian}, where $\nabla$ denotes the gradient operator. Another popular method is the wild bootstrap procedure; see, e.g., \cite{delgado2006consistent}, which requires a linear Bahadur expansion for the estimator of $\theta_0$. For the purpose of self-containment, we provide a brief overview of the orthogonality-based approach in Appendix \ref{sec:parameter_uncertainty}, along with the multiplier bootstrap procedure for approximating the null distribution of the test statistic. The asymptotic results established in previous sections remain valid after accounting for the estimation effect, a detailed discussion of which can also be found in Appendix \ref{sec:parameter_uncertainty}.

Our proposed test statistic now becomes 
$$ 
    n \hat{T}_{n,J(n)} = \sum_{j=1}^{J(n)} \left( \frac{1}{\sqrt{n}} \boldsymbol{\varepsilon}_{\hat{\theta}}^\top \hat{\boldsymbol{\Pi}}^\top \hat{e}_j (\boldsymbol{x}) \right)^2,
$$
where $\hat{\boldsymbol{\Pi}}$ is the sample version of the orthogonality operator $\boldsymbol{\Pi}$, refer to Appendix \ref{sec:parameter_uncertainty} for details. Throughout this section, we set the truncation level at $J(n) = 1$.  

\subsection{Simulation Studies}
\label{sec:simulation_study}
As benchmarks, we consider two KCM-type test statistics: the Gaussian process (GP) test statistic $\hat{T}_{GP}$ proposed by \cite{escanciano2024gaussian}, and the integrated conditional moment (ICM) test statistic $\hat{T}_{ICM}$ introduced by \cite{bierens1997asymptotic}. Both statistics are of the form\footnote{The classical ICM test statistic is constructed using the empirical process: $n \hat{T}_{n,ICM} = \int_\mathcal{X} |R_{n,exp}(t)|^2 N(dt)$, where $R_{n,exp}(t) = n^{-1/2} \sum_{\alpha=1}^n \varepsilon_{\hat \theta,\alpha} \exp(i x_\alpha^\top t)$, and $N$ is the standard normal distribution. However, it can be shown that, after considering the orthogonality operator, this statistic is numerically equivalent to the kernel-based form presented here when using the Gaussian kernel with bandwidth $\gamma = 1/2$.}:
\[
    n \hat T_{n,p} = \frac{1}{n} \boldsymbol{\varepsilon}_{\hat \theta}^\top \hat{\boldsymbol{\Pi}}^\top  \boldsymbol{K} \hat{\boldsymbol{\Pi}} \boldsymbol{\varepsilon}_{\hat \theta},  \, p \in \{GP, ICM\},
\]
where $\boldsymbol{K}$ is the kernel matrix with entries defined by $K_{\alpha,\beta} = k(x_\alpha, x_\beta)$ for $\alpha,\beta = 1,\ldots,n$. The GP test employs the Gaussian kernel $k(x,x') = \exp(-\gamma\|x-x'\|^2)$ with bandwidth $\gamma$ selected via the median heuristic $\gamma = 1 / \mathrm{median}(\{\|x_\alpha - x_\beta\| : \alpha \neq \beta\})$, while the ICM test also uses the Gaussian kernel but sets the bandwidth $\gamma = 1/2$.

We consider the following simulation designs. The null data-generating process (DGP) is specified as
\[
    \mathrm{DGP}_1: \quad Y = \zeta  + X^\top \xi  + e,
\]
where $X$ is a $d \times 1$ vector of independent standard normal variables, $e$ is an independent standard normal error, $\xi$ is a $d \times 1$ vector with all entries equal to $0.5$, and $\zeta = 1$. We examine two settings for the covariate dimension: $d = 10$ and $d = 20$.

The following fixed alternative models under homoskedasticity are considered:
\begin{align*}
    &\mathrm{DGP}_2:  \quad Y = \zeta + X^\top \xi + 1.5 \exp\left(- (X^\top \xi)^2 \right) + e,\\
    &\mathrm{DGP}_3:  \quad Y = \zeta + X^\top \xi + 2.0 \cos\left(1.1 \sqrt{X^\top X} \right) + e,\\
    &\mathrm{DGP}_4:  \quad Y = \zeta + X^\top \xi + 0.5 (X^\top \xi)^2 + e,\\
    &\mathrm{DGP}_5:  \quad Y = \zeta + X^\top \xi + 1.5 \exp\left(0.25 (X^\top \xi) \right) + e.
\end{align*}

For all DGPs $1$--$5$, the covariate and error structures are as described above.

To assess performance under heteroskedasticity, we consider:
\begin{align*}
    &\mathrm{DGP}_6: \quad Y = \zeta + X^\top \xi + \sqrt{X^\top X} + c_1(X) e, \quad \text{for } d=10,\\
    &\mathrm{DGP}_6^{\ast}: \quad Y = \zeta + X^\top \xi + \sqrt{X^\top X} + c_2(X) e, \quad \text{for } d=20.
\end{align*}
For $d=10$, the first five covariates $\{X_{l}\}_{l=1}^{5}$ are drawn independently from $\mathrm{Uniform}[0,\, 1+0.1(l-1)]$, and the remaining $\{X_{l}\}_{l=6}^{10}$ from $\mathrm{Normal}(0,\, 1+0.1(l-5))$. For $d=20$, $\{X_{l}\}_{l=1}^{10}$ are drawn from $\mathrm{Uniform}[0,\, 1+0.1(l-1)]$, and $\{X_{l}\}_{l=11}^{20}$ from $\mathrm{Normal}(0,\, 1+0.1(l-10))$. The heteroskedasticity functions are $c_1(X) = |X^\top \boldsymbol{1}|$ and $c_2(X) = |X^\top \boldsymbol{1}|$, where $\boldsymbol{1}$ is a $d \times 1$ vector of ones. In both cases, $\xi = \boldsymbol{1}$, $\zeta = 1$, and $e$ is standard normal.

For local alternatives, we consider:
\begin{align*}
    &\mathrm{DGP}_7: \quad Y = \zeta + X^\top \xi + \frac{\sqrt{X^\top X}}{\sqrt{n}} + d_1(X) e, \quad \text{for } d=10,\\
    &\mathrm{DGP}_7^{\ast}: \quad Y = \zeta + X^\top \xi + \frac{\sqrt{X^\top X}}{\sqrt{n}} + d_2(X) e, \quad \text{for } d=20.
\end{align*}
For $d=10$, $\{X_{l}\}_{l=1}^{5}$ are drawn from $\mathrm{Uniform}[0,\, l]$, and $\{X_{l}\}_{l=6}^{10}$ from $\mathrm{Normal}(0,\, 1+0.1(l-5))$. For $d=20$, $\{X_{l}\}_{l=1}^{10}$ are drawn from $\mathrm{Uniform}[0,\, l]$, and $\{X_{l}\}_{l=11}^{20}$ from $\mathrm{Normal}(0,\, 1+0.1(l-10))$. The heteroskedasticity functions are $d_1(X) = \sqrt{0.1 + \sum_{l=1}^{5} X_{l} + \sum_{l=6}^{10} X_{l}^2}$ and $d_2(X) = \sqrt{0.1 + \sum_{l=1}^{5} X_{l} + \sum_{l=6}^{20} X_{l}^2}$. In both cases, $\xi$ is a vector of ones, $\zeta = 1$, and $e$ is standard normal.

The simulation results are based on $1000$ replications, and in each replication, we set the bootstrap sample size to $B = 500$. The nominal significance levels are set to $10\%,5\%$ and $1\%$, and the critical values are obtained via the multiplier bootstrap method described in Appendix~\ref{sec:bootstrap}. 

The comparison results between the truncated and non-truncated test statistics with pre-determined kernel are presented in Tables~\ref{tab:truncate_all_comparison_q10} and \ref{tab:truncate_all_comparison_q20}. Here, we focus on the benchmark test statistics $\hat{T}_{n,GP}$ and $\hat{T}_{n,ICM}$, as well as their truncated versions $\hat{T}_{n,J(n),GP}$ and $\hat{T}_{n,J(n),ICM}$. We draw several conclusions from these results. First, the truncated test statistics generally exhibit better size control than their non-truncated counterparts, particularly in higher dimensions. Second, the power of the truncated tests is also better in most scenarios, especially under high-dimensional settings. Third, the performance of GP tests, which use the median heuristic for bandwidth selection, is superior to that of ICM tests with fixed bandwidth, particularly in higher dimensions. The last observation highlights the importance of selecting an appropriate bandwidth in kernel-based tests, a topic that warrants further research.

It is important to emphasize that the truncated test statistic is not universally superior to its non-truncated counterpart. ICM-based tests are well-established as admissible, meaning no other test uniformly dominates them in terms of power. Instead, the truncation strategy serves as a practical method to enhance the signal-to-noise ratio in finite samples. The effectiveness of this approach depends on the truncation level $J(n)$, which is generally challenging to optimize without additional prior knowledge.

\newlength{\colwidth}
\setlength{\colwidth}{0.085\textwidth} % Adjust as needed

\begin{table}[htbp]
    \centering
    \caption{Truncated and Non-Truncated Test Statistics with Fixed Kernel at Multiple Significance Levels ($\alpha = 10\%, 5\%, 1\%$) and $d=10$}
    {\small % Reduce font size for the table
    \begin{tabular}{
        @{} 
        >{\raggedright\arraybackslash}p{0.12\textwidth} % First column (left-aligned, slightly wider)
        *{8}{>{\centering\arraybackslash}p{\colwidth}} % Remaining columns (centered)
        @{}
    }
        \toprule
        \multicolumn{1}{@{}>{\raggedright\arraybackslash}p{0.12\textwidth}}{} 
        & \multicolumn{4}{c}{$n = 200$} 
        & \multicolumn{4}{c}{$n = 400$} \\
        \cmidrule(lr){2-5} \cmidrule(lr){6-9}
        & $\hat{T}_{n,GP}$ & $\hat{T}_{n,J(n),GP}$ & $\hat{T}_{n,ICM}$ & $\hat{T}_{n,J(n),ICM}$ & $\hat{T}_{n,GP}$ & $\hat{T}_{n,J(n),GP}$ & $\hat{T}_{n,ICM}$ & $\hat{T}_{n,J(n),ICM}$ \\
        \midrule
        % SIZE Section
        \multicolumn{9}{c}{SIZE ($DGP_1$)} \\
        \midrule
        $\alpha = 10\%$ & 0.041 & 0.095 & 0.000 & 0.105 & 0.047 & 0.110 & 0.000 & 0.103 \\
        $\alpha = 5\%$  & 0.008 & 0.042 & 0.000 & 0.063 & 0.017 & 0.051 & 0.000 & 0.050 \\
        $\alpha = 1\%$  & 0.000 & 0.009 & 0.000 & 0.011 & 0.001 & 0.008 & 0.000 & 0.010 \\
        \midrule
        % POWER Section
        \multicolumn{9}{c}{POWER} \\
        \midrule
        \multicolumn{9}{c}{$DGP_2$} \\
        \midrule
        $\alpha = 10\%$ & 0.544 & 0.435 & 0.022 & 0.403 & 0.945 & 0.665 & 0.277 & 0.651 \\
        $\alpha = 5\%$  & 0.273 & 0.311 & 0.001 & 0.288 & 0.869 & 0.558 & 0.064 & 0.523 \\
        $\alpha = 1\%$  & 0.046 & 0.130 & 0.000 & 0.113 & 0.550 & 0.318 & 0.000 & 0.274 \\
        \midrule
        \multicolumn{9}{c}{$DGP_3$} \\
        \midrule
        $\alpha = 10\%$ & 0.163 & 0.371 & 0.008 & 0.125 & 0.394 & 0.578 & 0.076 & 0.117 \\
        $\alpha = 5\%$  & 0.068 & 0.283 & 0.000 & 0.064 & 0.222 & 0.450 & 0.009 & 0.060 \\
        $\alpha = 1\%$  & 0.008 & 0.119 & 0.000 & 0.014 & 0.053 & 0.259 & 0.000 & 0.009 \\
        \midrule
        \multicolumn{9}{c}{$DGP_4$} \\
        \midrule
        $\alpha = 10\%$ & 0.965 & 0.983 & 0.085 & 0.953 & 1.000 & 1.000 & 0.766 & 0.998 \\
        $\alpha = 5\%$  & 0.892 & 0.963 & 0.006 & 0.909 & 0.999 & 1.000 & 0.463 & 0.996 \\
        $\alpha = 1\%$  & 0.552 & 0.850 & 0.000 & 0.768 & 0.996 & 1.000 & 0.046 & 0.987 \\
        \midrule
        \multicolumn{9}{c}{$DGP_5$} \\
        \midrule
        $\alpha = 10\%$ & 0.065 & 0.196 & 0.003 & 0.143 & 0.167 & 0.267 & 0.001 & 0.205 \\
        $\alpha = 5\%$  & 0.024 & 0.120 & 0.000 & 0.084 & 0.077 & 0.171 & 0.000 & 0.132 \\
        $\alpha = 1\%$  & 0.000 & 0.036 & 0.000 & 0.024 & 0.012 & 0.050 & 0.000 & 0.032 \\
        \midrule
        \multicolumn{9}{c}{$DGP_6$} \\
        \midrule
        $\alpha = 10\%$ & 0.384 & 0.774 & 0.094 & 0.756 & 0.756 & 0.969 & 0.355 & 0.947 \\
        $\alpha = 5\%$  & 0.244 & 0.689 & 0.030 & 0.655 & 0.629 & 0.926 & 0.180 & 0.898 \\
        $\alpha = 1\%$  & 0.083 & 0.441 & 0.000 & 0.402 & 0.361 & 0.807 & 0.032 & 0.748 \\
        \midrule
        \multicolumn{9}{c}{$DGP_7$} \\
        \midrule
        $\alpha = 10\%$ & 0.095 & 0.393 & 0.000 & 0.256 & 0.128 & 0.372 & 0.000 & 0.306 \\
        $\alpha = 5\%$  & 0.031 & 0.273 & 0.000 & 0.157 & 0.046 & 0.249 & 0.000 & 0.220 \\
        $\alpha = 1\%$  & 0.001 & 0.102 & 0.000 & 0.048 & 0.006 & 0.109 & 0.000 & 0.076 \\
        \bottomrule
    \end{tabular}
    } % End of \small scope
    \label{tab:truncate_all_comparison_q10}
\end{table}

\begin{table}[htbp]
    \centering
    \caption{Truncated and Non-Truncated Test Statistics with Fixed Kernel at Multiple Significance Levels ($\alpha = 10\%, 5\%, 1\%$) and $d = 20$}
    {\small % Reduce font size for the table
    \begin{tabular}{
        @{} 
        >{\raggedright\arraybackslash}p{0.12\textwidth} % Slightly wider first column for significance labels
        *{8}{>{\centering\arraybackslash}p{0.085\textwidth}} 
        @{}
    }
        \toprule
        \multicolumn{1}{@{}>{\raggedright\arraybackslash}p{0.12\textwidth}}{} 
        & \multicolumn{4}{c}{$n = 200$} 
        & \multicolumn{4}{c}{$n = 400$} \\
        \cmidrule(lr){2-5} \cmidrule(lr){6-9}
        & $\hat{T}_{n,GP}$ & $\hat{T}_{n,J(n),GP}$ & $\hat{T}_{n,ICM}$ & $\hat{T}_{n,J(n),ICM}$ & $\hat{T}_{n,GP}$ & $\hat{T}_{n,J(n),GP}$ & $\hat{T}_{n,ICM}$ & $\hat{T}_{n,J(n),ICM}$ \\
        \midrule
        % SIZE Section
        \multicolumn{9}{c}{SIZE ($DGP_1$)} \\
        \midrule
        $\alpha = 10\%$ & 0.012 & 0.135 & 0.000 & 0.094 & 0.002 & 0.121 & 0.000 & 0.094 \\
        $\alpha = 5\%$  & 0.000 & 0.073 & 0.000 & 0.023 & 0.000 & 0.055 & 0.000 & 0.025 \\
        $\alpha = 1\%$  & 0.000 & 0.022 & 0.000 & 0.001 & 0.000 & 0.012 & 0.000 & 0.001 \\
        \midrule
        % POWER Section
        \multicolumn{9}{c}{POWER} \\
        \midrule
        \multicolumn{9}{c}{$DGP_2$} \\
        \midrule
        $\alpha = 10\%$ & 0.031 & 0.211 & 0.000 & 0.131 & 0.026 & 0.343 & 0.000 & 0.154 \\
        $\alpha = 5\%$  & 0.001 & 0.127 & 0.000 & 0.033 & 0.000 & 0.240 & 0.000 & 0.051 \\
        $\alpha = 1\%$  & 0.000 & 0.034 & 0.000 & 0.001 & 0.000 & 0.092 & 0.000 & 0.008 \\
        \midrule
        \multicolumn{9}{c}{$DGP_3$} \\
        \midrule
        $\alpha = 10\%$ & 0.309 & 0.749 & 0.001 & 0.173 & 0.429 & 0.955 & 0.000 & 0.232 \\
        $\alpha = 5\%$  & 0.031 & 0.663 & 0.000 & 0.061 & 0.143 & 0.905 & 0.000 & 0.093 \\
        $\alpha = 1\%$  & 0.000 & 0.460 & 0.000 & 0.018 & 0.007 & 0.758 & 0.000 & 0.031 \\
        \midrule
        \multicolumn{9}{c}{$DGP_4$} \\
        \midrule
        $\alpha = 10\%$ & 0.069 & 0.917 & 0.000 & 0.365 & 0.487 & 0.999 & 0.000 & 0.600 \\
        $\alpha = 5\%$  & 0.001 & 0.854 & 0.000 & 0.199 & 0.096 & 0.994 & 0.000 & 0.435 \\
        $\alpha = 1\%$  & 0.000 & 0.654 & 0.000 & 0.042 & 0.002 & 0.951 & 0.000 & 0.204 \\
        \midrule
        \multicolumn{9}{c}{$DGP_5$} \\
        \midrule
        $\alpha = 10\%$ & 0.039 & 0.272 & 0.000 & 0.121 & 0.017 & 0.410 & 0.000 & 0.163 \\
        $\alpha = 5\%$  & 0.000 & 0.172 & 0.000 & 0.043 & 0.001 & 0.291 & 0.000 & 0.054 \\
        $\alpha = 1\%$  & 0.000 & 0.057 & 0.000 & 0.003 & 0.000 & 0.108 & 0.000 & 0.004 \\
        \midrule
        \multicolumn{9}{c}{$DGP_6$} \\
        \midrule
        $\alpha = 10\%$ & 0.026 & 0.439 & 0.000 & 0.110 & 0.039 & 0.658 & 0.000 & 0.150 \\
        $\alpha = 5\%$  & 0.000 & 0.312 & 0.000 & 0.033 & 0.003 & 0.526 & 0.000 & 0.053 \\
        $\alpha = 1\%$  & 0.000 & 0.130 & 0.000 & 0.000 & 0.000 & 0.284 & 0.000 & 0.007 \\
        \midrule
        \multicolumn{9}{c}{$DGP_7$} \\
        \midrule
        $\alpha = 10\%$ & 0.000 & 0.258 & 0.000 & 0.088 & 0.000 & 0.285 & 0.000 & 0.073 \\
        $\alpha = 5\%$  & 0.000 & 0.177 & 0.000 & 0.001 & 0.000 & 0.191 & 0.000 & 0.000 \\
        $\alpha = 1\%$  & 0.000 & 0.063 & 0.000 & 0.000 & 0.000 & 0.068 & 0.000 & 0.000 \\
        \bottomrule
    \end{tabular}
    } % End of \small scope
    \label{tab:truncate_all_comparison_q20}
\end{table}

\subsection{Empirical Applications}
\label{sec:empirical_application}
In this section, we demonstrate the practical utility of our proposed test statistics by applying them to two real-world empirical settings. 

\textit{Linear Regression Models}. In the first application, we revisit the linear regression analysis of \cite{gazeaud2023or}, which addresses two key questions: (i) Can relaxing financial capital constraints stimulate women's income-generating activities (IGAs)? and (ii) Does involving husbands in these interventions enhance or impede their effectiveness? The dataset used in this analysis is publicly available at \url{https://doi.org/10.7910/DVN/VDB0YJ}.

The authors conducted a three-arm randomized controlled trial (RCT) to evaluate the impact of a cash grant and gender-sensitive training program for women in rural Tunisia. The study included two treatment arms. In the first arm, 1,000 women received an unrestricted cash grant equivalent to USD 768 (in 2018 PPP terms)—a substantial sum, representing approximately four times the median monthly income of participants at baseline. Alongside the grant, these women attended a one-day financial training session featuring videos and exercises designed to promote women's agency. Participants were encouraged to invest the grant in IGAs or in human capital to improve their labor market prospects.

The second treatment arm sought to address gender-specific barriers by involving husbands in the intervention. Specifically, 498 of the 1,000 women who received the cash grant were randomly selected to invite their husbands to participate in the same one-day training, which aimed to foster gender dialogue and support for women's economic activities.   

The impacts of the treatments were evaluated two years after implementation. The authors categorize the outcomes into several domains, including women's \textit{income-generating activities} (IGAs), \textit{empowerment}, \textit{well-being}, \textit{living standards}, and \textit{economic shocks}. Their empirical strategy is based on the following regression specification:
\[
    y_\alpha = \zeta + \zeta_1 T_{\alpha 1} + \zeta_2 T_{\alpha 2} + \xi^\top X_\alpha + \varepsilon_\alpha, \quad \alpha = 1, \ldots, n,
\]
where $y_\alpha$ denotes the outcome of interest for individual $\alpha$, $T_{\alpha 1}$ and $T_{\alpha 2}$ are binary indicators for assignment to the two treatment arms (cash grant only and cash grant with husband training, respectively), $X_\alpha$ is a vector of control variables, and $\varepsilon_\alpha$ is the error term. The intent-to-treat (ITT) effects are captured by the coefficients $\zeta_1$ and $\zeta_2$. To address the second research question—-whether involving husbands enhances or diminishes the effectiveness of the intervention—-the authors test the null hypothesis $\zeta_1 = \zeta_2$. In their complete specification model, the authors include 30 control variables. We refer to their paper for details on the data and the control variables. The final sample consists of $n = 1,824$ observations.

The authors examine 37 outcomes, but we focus on five key measures: (i) \textit{IGAs}, a binary indicator equal to one if the woman engaged in an income-generating activity at the time of the survey; (ii) \textit{financial index}, the standardized average of nine questions assessing women's financial access and situation, used as a proxy for empowerment; (iii) \textit{mental health}, measured by the MHI-5 score, which aggregates responses on happiness, peacefulness, nervousness, downheartedness, and depression (ranging from 0 to 100, with higher values indicating better mental health); (iv) \textit{asset index}, a standardized index based on ownership of 21 assets; and (v) \textit{economic shocks}, a binary indicator equal to one if the respondent experienced job loss, business failure, or loss of livelihood in the past 24 months.

Among these outcomes, the first and last are binary indicators, while the others are continuous variables. Assessing the adequacy of the linear regression model is particularly important for the binary outcomes, where the linearity assumption may be violated.

Table~\ref{tab:empirical_results} presents the multiplier bootstrap $p$-values for the five key outcomes using four test statistics: the non-truncated GP and ICM tests ($\hat{T}_{n,GP}$ and $\hat{T}_{n,ICM}$) and their truncated counterparts ($\hat{T}_{n,J(n),GP}$ and $\hat{T}_{n,J(n),ICM}$). For most outcomes, the $p$-values are well above conventional significance levels, indicating no substantial evidence against the linear model specification. However, the truncated ICM test yields a $p$-value of 0.072 for the \textit{Mental health} outcome, suggesting potential model misspecification. Similarly, the truncated GP test produces a $p$-value of 0.032 for the \textit{Economic shocks} outcome, pointing to possible misspecification in this case as well. 

\setlength{\colwidth}{0.15\textwidth}
\begin{table}[ht]
\centering
\caption{Multiplier Bootstrap $p$-value Results for Different Outcomes and Different Test Statistics}
\begin{tabular}{
    @{}
    >{\raggedright\arraybackslash}p{0.2\textwidth}
    *{5}{>{\centering\arraybackslash}p{\colwidth}}
    @{}
}
\toprule
 & $\hat{T}_{n,GP}$ & $\hat{T}_{n,J(n),GP}$ & $\hat{T}_{n,ICM}$ & $\hat{T}_{n,J(n),ICM}$ \\
\midrule
IGA             & 0.498 & 0.330 & 0.518 & 0.294 \\
Financial index & 0.492 & 0.158 & 0.466 & 0.242 \\
Mental health   & 0.400 & 0.748 & 0.504 & 0.072 \\
Asset index     & 0.198 & 0.184 & 0.470 & 0.282 \\
Economic shocks & 0.432 & 0.032 & 0.502 & 0.278 \\
\bottomrule
\end{tabular}
\label{tab:empirical_results}
\end{table}

\textit{Probit Regression}. In the second application, we assess the adequacy of a probit regression model employed by \cite{hu2023long}.\footnote{The data can be obtained on request from the original authors.} The authors examine how foreign aid programs and their legacies influence international relations between countries. Specifically, they provide empirical evidence on the long-term effects of such aid on international trade by analyzing the 156 Program in China, which was supported by the Soviet Union. 

The 156 Program was a large-scale industrial aid initiative launched in the 1950s, which involved the construction of 156 major industrial enterprises across China. Its primary objective was to strengthen China's industrial capacity and promote economic self-sufficiency, with far-reaching consequences for both domestic development and international relations. However, the locations of these aid projects were not randomly assigned. Instead, they were selected based on explicit criteria: (i) proximity to resources and railway networks, (ii) placement in underdeveloped regions, and (iii) locations beyond the reach of potential American air raids. This non-random assignment of project sites raises significant concerns regarding the identification of the causal effect of aid on international trade. 

To address this issue, the authors employ a probit regression model to estimate the propensity scores for each city, thereby matching recipient and control cities with similar characteristics. The explanatory variables reflecting the first criterion include the distance to railways in 1949 and travel distances to coal and iron ore mines in 1953. To proxy for the level of economic development, they use population density, the proportion of rural residents, and the average years of education in 1953. The third criterion—exposure to potential American air raids—is captured by the average distance to the nearest American airbase.

To assess the adequacy of the probit model, we apply the test statistics $\hat{T}_{n,GP}$, $\hat{T}_{n,J(n),GP}$, $\hat{T}_{n,ICM}$, and $\hat{T}_{n,J(n),ICM}$, , with the resulting multiplier bootstrap $p$-values reported in Table~\ref{tab:probit_results}. The analysis is based on a sample of $N = 336$ cities, where the outcome variable is a binary indicator equal to one if the city received aid under the 156 Program. All test statistics yield $p$-values well above conventional significance thresholds, providing no evidence against the probit model specification. These results suggest that the probit model is suitable for the data, thereby supporting the validity of the authors' propensity score estimation approach.

\setlength{\colwidth}{0.15\textwidth}
\begin{table}[ht]
\centering
\caption{Multiplier Bootstrap $p$-value Results for the Probit Model for Different Test Statistics}
\begin{tabular}{
    @{}
    >{\raggedright\arraybackslash}p{0.2\textwidth}
    *{5}{>{\centering\arraybackslash}p{\colwidth}}
    @{}
}
\toprule
 & $\hat{T}_{n,GP}$ & $\hat{T}_{n,J(n),GP}$ & $\hat{T}_{n,ICM}$ & $\hat{T}_{n,J(n),ICM}$ \\
\midrule
$p$-values   & 0.358 & 0.692 & 0.294 & 0.748 \\
\bottomrule
\end{tabular}
\label{tab:probit_results}
\end{table}

\section{Conclusion}
\label{sec:conclusion}
This paper examines the discrepancy between the empirical and population versions of the directional components within the KCM test statistic. The primary tools used in our analysis are perturbation theory and recent results on the convergence of eigenspaces. The main messages of our work are as follows:

\begin{enumerate}
    \item For a fixed sample size, the magnitude of the discrepancy varies across directional components, with high-frequency components (those associated with small eigenvalues) being more susceptible to estimation error. This implies that empirical directional components corresponding to small eigenvalues are unlikely to capture the true deviation from the null hypothesis accurately.
    
    \item Consequently, truncating high-frequency components can enhance the finite-sample performance of the KCM test. The concern that such truncation may lead to power loss--due to discarding potentially informative directions--is mitigated by the fact that these high-frequency components are poorly estimated and thus contribute little to the test’s power. Moreover, removing them reduces the noise they introduce, thereby improving the overall signal-to-noise ratio of the test statistic.
    
    \item There exists a trade-off in the choice of the truncation level: a higher truncation level retains more components, potentially increasing power, but also incorporates more noise due to estimation error in the low-eigenvalue directions. This trade-off is strongly influenced by the sample size, the decay rate of the kernel eigenvalues, and the nature of the alternative hypothesis.
\end{enumerate}  
Future works may explore the following directions: (i) the optimal selection of the truncation level in a data-driven manner; (ii) the choice of kernel (and equivalently, the eigenvalue decay rate) to maximize test power; (iii) the extension of our analysis to other kernel-based tests, such as those based on Maximum Mean Discrepancy (MMD) or Hilbert--Schmidt Independence Criterion (HSIC).

\section*{Funding Sources}
This work was supported by the Research Development Fund (RDF-23-02-022) of the Xi'an Jiaotong--Liverpool University and the National Natural Science Foundation of China (Grant Numbers 72373007 and 72333001).

\bibliography{paper-ref}

\newpage
\appendix
\begin{center}
    \Large\textbf{Online Appendix}
\end{center}
\section{Useful Lemmas and Theorems}
\label{sec:useful_lemmas}
\begin{lemma}
    \label{lemma:upper_bound_operator}
    On a Hilbert space $\mathcal{H}$, let $\mathcal{T}$ and $\mathcal{\hat{T}}$ be nonnegative definite, compact operators with eigenvalues $\{\lambda_j\}_{j=1}^\infty$ and $\{\hat{\lambda}_j\}_{j=1}^\infty$, respectively. Then
    $$ 
        \sup_{j \geq 1} |\lambda_j - \hat{\lambda}_j| \leq ||\mathcal{T} - \mathcal{\hat{T}}||_{op},
    $$  
    where $||\cdot||_{op}$ is the operator norm.
\end{lemma}
\begin{proof}
    See Theorem 4.2.8 in \cite{hsing2015theoretical}.
\end{proof}

\begin{lemma}
    \label{lemma:HS_bound}
    Suppose that $\sup_{x \in \mathcal{X}}k(x,x) \leq \kappa$. Then, with probability at least $1-e^{-\xi}$, we have 
    $$ 
        \| \Sigma_{n} - \Sigma \|_{HS} \leq \frac{2\kappa}{\sqrt{n}}\left(1+ \sqrt{\frac{\xi}{2}}\right),
    $$  
    where $\Sigma$ and $\Sigma_n$ are the population and empirical (uncentered) covariance operators, respectively, defined in Section~\ref{sec:RKHS_intro}.
\end{lemma}
\begin{proof}
    See \cite{zwald2005convergence}.
\end{proof}
Since both $\Sigma$ and $\Sigma_n$ positive definite and compact, and 
$$ 
    \| \cdot \|_{op} \leq \| \cdot \|_{HS},
$$
applying Lemma~\ref{lemma:upper_bound_operator} leads to:
$$ 
     |\lambda_j - \hat{\lambda}_j| = O_p(1/\sqrt{n}), \quad \forall j \geq 1.
$$

\section{Parameter Uncertainty and Multiplier Bootstrap}
\label{sec:parameter_uncertainty}
\subsection{Parameter Uncertainty}
We follow the orthogonality-based approach proposed by \cite{escanciano2014specification,sant2019specification,escanciano2024gaussian}, which uses Neyman orthogonality to mitigate the impact of parameter estimation on the test statistic. Unlike traditional wild bootstrap procedures (see, e.g., \cite{delgado2006consistent}), the orthogonality-based strategy does not rely on a linear Bahadur expansion for the estimator of $\theta_0$, nor does it require re-estimating parameters in each bootstrap replication.

The following assumptions are imposed to facilitate this approach.
\begin{assumption}
    The consistent estimator $\hat{\theta}$ satisfies $\lVert \hat{\theta} - \theta_0 \rVert = O_p (n^{-\beta})$, with $\beta > 1/4$.
\end{assumption}
This condition is weaker than the related ones in the literature. We only require that $\hat{\theta}$ converges in probability, potentially at a slower rate than is typical. Moreover, we do not assume that $\hat{\theta}$ admits an asymptotically linear representation. This flexibility is particularly useful when considering non-standard estimation procedures, such as the LASSO.

Additional regular conditions on the smoothness of the residual function are also required.
\begin{assumption}
    (i) The residual $\varepsilon_{\theta}$ is twice continuously differentiable with respect to $\theta$, with its first derivative $g_{\theta}(x) = \mathbb{E}(\nabla_{\theta} \varepsilon_{\theta}|X=x)$ satisfying $\mathbb{E}\left(\sup_{\theta \in \Theta} \lVert g_{\theta}(X) \rVert\right) < \infty$ and its second derivative satisfying $\mathbb{E}\left(\sup_{\theta \in \Theta} \lVert \nabla g_{\theta}(X) \rVert\right) < \infty$; (ii) the matrix $\Gamma_{\theta} = \mathbb{E}\left[g_{\theta}(X) g_{\theta}(X) ^\top\right]$ is nonsingular in a neighborhood of $\theta_0$.
\end{assumption}

Under these assumptions, we can now introduce an orthogonal operator $\boldsymbol{\Pi}$, acting on the feature map $k(t,\cdot)$ indexed by $t \in \mathcal{X}$, defined as:
$$ 
\boldsymbol{\Pi}k(t,x) = k(t,x) -(g_{\theta_0}(x))^\top  \Gamma_{\theta_0}^{-1} \mathbb{E}\left[k(t,X) g_{\theta_0}(X) \right], \forall x \in \mathcal{X}.
$$   

To analyze the local behavior of these modified directional components near $\theta_0$, consider their derivatives with respect to $\theta$ evaluated at $\theta_0$:
\begin{align*}
    \nabla_{\theta = \theta_0} \mathbb{E}\left(\varepsilon_{\theta} \boldsymbol{\Pi} k(t,X)\right) 
    &= \mathbb{E} \left[ g_{\theta_0}(X) k(t,X) - (g_{\theta_0}(X))^\top g_{\theta_0}(X) \Gamma_{\theta_0}^{-1} \mathbb{E}\left(k(t,X) g_{\theta_0}(X) \right) \right] \\
    &= \boldsymbol{0}.
\end{align*}
This result shows that the modified directional components are Neyman orthogonal at $\theta_0$, meaning they are locally insensitive to small perturbations in $\theta$. As a consequence, the impact of estimating $\theta_0$ on the test statistic is asymptotically negligible, ensuring the validity of inference even when $\theta_0$ is replaced by a consistent estimator $\hat{\theta}$. 

The matrix estimator (using the test data) of this projection operator is given by:
$$ 
    \hat{\boldsymbol{\Pi}} = \boldsymbol{I}_{n} -  \mathbb{G}\left(\mathbb{G}^\top \mathbb{G}\right)^{-1} \mathbb{G}^\top
$$
where $\mathbb{G}$ is a $n \times d$ matrix of scores whose $i$th row is given by $\hat g_{i}^\top = (\nabla_{\theta} \varepsilon_{\theta,i}|_{\theta = \hat \theta})^\top$, and $\boldsymbol{I}_{n}$ is the $n \times n$ identity matrix.

The following lemma states how this projection operator eliminates the estimation effect in finite samples. 

\begin{lemma}
    \label{lem:projection_finite_sample}
    Suppose the Assumptions stated in this section hold, then 
    \[
        \frac{1}{\sqrt{n}}  (\hat{\boldsymbol{\Pi}} \boldsymbol{\varepsilon}_{\hat \theta})^\top \hat{e}_j(\boldsymbol{x})  = \frac{1}{\sqrt{n}} (\hat{\boldsymbol{\Pi}} \boldsymbol{\varepsilon}_{\theta_0})^\top \hat{e}_j(\boldsymbol{x}) + o_p(1).
    \]

    In particular, when the null hypothesis holds, we have
    $$ 
        \frac{1}{\sqrt{n}}  (\hat{\boldsymbol{\Pi}} \boldsymbol{\varepsilon}_{\hat \theta})^\top \hat{e}_j(\boldsymbol{x})  = \frac{1}{\sqrt{n}} (\hat{\boldsymbol{\Pi}} \boldsymbol{\varepsilon}_{\theta_0})^\top e_j(\boldsymbol{x}) + o_p(1).
    $$
\end{lemma}

\begin{proof}
    For the first statement, see Appendix~\ref{sec:proof_projection_finite_sample}. The second statement follows directly from Theorem \ref{thm:residual_asymptotics} in the main text.
\end{proof}

\subsection{Multiplier Bootstrap}   
\label{sec:bootstrap}
We use the multiplier bootstrap to approximate the null distributions of the proposed test statistics. For theoretical justification, we adopt the notion of almost sure (a.s.) consistency, denoted by $\overset{d^*}{\longrightarrow}$; see Chapter 2.9 of \cite{vaart1997weak}. The procedure involves generating a sequence of i.i.d. random variables $\{v_\alpha\}_{\alpha=1}^{n}$ that are independent of the data $\{y_\alpha,x_\alpha\}_{\alpha=1}^{n}$, have zero mean, unit variance, and bounded support. These multipliers are then used to construct the bootstrap sample $\{\varepsilon_{\hat \theta,\alpha} v_\alpha\}_{\alpha=1}^{n}$. A common choice for the multipliers is Mammen's two-point distribution:
\[
    \mathbb{P}(V_\alpha = 0.5(1-\sqrt{5})) = b, \quad \mathbb{P}(V_\alpha = 0.5(1+\sqrt{5})) = 1-b,
\]
where $b = (1+\sqrt{5})/(2\sqrt{5})$; see \cite{mammen1993bootstrap} for details.

The bootstrap analogue of 
$$ 
    \sum_{j=1}^{J(n)}\left(\frac{1}{\sqrt{n}}  (\hat{\boldsymbol{\Pi}} \boldsymbol{\varepsilon}_{\hat \theta})^\top \hat{e}_j(\boldsymbol{x})\right)^2
$$
is given by
$$ 
    \sum_{j=1}^{J(n)} \left(\frac{1}{\sqrt{n}}(\hat{\boldsymbol{\Pi}} (\boldsymbol{\varepsilon}_{\hat \theta}\odot \boldsymbol{V}))^\top \hat{e}_j(\boldsymbol{x})\right)^2,
$$
where $\odot$ denotes the Hadamard product, and $\boldsymbol{V} = (v_1,\ldots,v_n)^\top$. 

\begin{theorem}
    \label{thm:bootstrap}
    Under Assumptions stated in this section, as the sample sizes $n \to \infty$,  we have 
    $$ 
        \sum_{j=1}^{J(n)} \left(\frac{1}{\sqrt{n}}(\hat{\boldsymbol{\Pi}} (\boldsymbol{\varepsilon}_{\hat \theta}\odot \boldsymbol{V}))^\top \hat{e}_j(\boldsymbol{x})\right)^2 \overset{d^*}{\longrightarrow} \sum_{i=1}^{\infty} \lambda_j  W_j^2,
    $$
    with covariance structure $\mathrm{Cov}(W_j,W_i) = \mathbb{E}(\varepsilon_{\theta_0}^2\phi_j(X)\phi_i(X))$. 
\end{theorem} 
\begin{proof}
    See Appendix~\ref{sec:proof_bootstrap}.
\end{proof}

\section{Proofs}
\label{sec:proofs}
\subsection{Proof of Lemma \ref{lemma:analytic_perturbation}}
We study the term $\langle e_i, (\hat{\Sigma} - \Sigma) e_j \rangle_{\mathcal{H}_k}$. Note that
\begin{align*}
    \langle e_i, (\hat{\Sigma} - \Sigma) e_j \rangle_{\mathcal{H}_k} & = \langle e_i, \hat{\Sigma} e_j \rangle_{\mathcal{H}_k} - \langle e_i, \Sigma e_j \rangle_{\mathcal{H}_k} \\
    & = \langle e_i, \hat{\Sigma} e_j \rangle_{\mathcal{H}_k} \\
    & = \frac{1}{n} \sum_{\alpha=1}^n e_j(x_\alpha) \langle e_i, k(x_\alpha, \cdot) \rangle_{\mathcal{H}_k} \\
    & = \frac{1}{n} \sum_{\alpha=1}^n e_j(x_\alpha) e_i(x_\alpha)\\
    & = \frac{\sqrt{\lambda_i \lambda_j}}{n} \sum_{\alpha=1}^n \phi_i(x_\alpha) \phi_j(x_\alpha).
\end{align*} 
The second equality follows from the orthogonality of $e_j$ and $e_i$ in $\mathcal{H}_k$. The last equality follows from the reproducing property of RKHS. 

Note that 
$$ 
    \left|\sum_{i \neq j} \frac{\lambda_i}{\lambda_j - \lambda_i} 
    \right| \leq \sum_{i=1}^{\infty} \frac{\lambda_i}{2 \eta_j} < \infty,
$$
since $\sum_{i=1}^{\infty}\lambda_i < \infty$ by the compactness of $L_k$. The rest follows immediately by employing Theorem \ref{thm:upper_bound_eigenfunction} and simple algebraic manipulations.

\subsection{Proof of Theorem \ref{thm:residual_asymptotics}}
Using the results in Lemma~\ref{lemma:analytic_perturbation}, we have
\begin{align*}
    (\sqrt{n})^{-1} \boldsymbol{\varepsilon}_{\theta_0}^\top (\hat{e}_j - e_j)(\boldsymbol{x}) & = \sqrt{\lambda_j}(\sqrt{n})^{-1} \sum_{i \neq j} \frac{\lambda_i}{\lambda_j - \lambda_i} \left(n^{-1}\sum_{\alpha=1}^n \phi_i(x_\alpha)\phi_j(x_\alpha)\right) \boldsymbol{\varepsilon}_{\theta_0}^\top \phi_i(\boldsymbol{x}) + O_p(n^{-1/2}) \\
    & = \sqrt{\lambda_j}\sum_{i \neq j} \frac{\lambda_i}{\lambda_j - \lambda_i} \left((\sqrt{n})^{-1}\sum_{\alpha=1}^n \phi_i(x_\alpha)\phi_j(x_\alpha)\right) n^{-1} \boldsymbol{\varepsilon}_{\theta_0}^\top \phi_i(\boldsymbol{x}) + o_p(1).
\end{align*}
The term 
$$ 
    (\sqrt{n})^{-1}\sum_{\alpha=1}^n \phi_i(x_\alpha)\phi_j(x_\alpha) \overset{d}{\longrightarrow} W_{ij}, 
$$
since by the orthogonality of eigenfunctions, $\mathbb{E}(\phi_i(X)\phi_j(X)) = 0$ for $i \neq j$, and $W_{ij} \sim \mathcal{N}(0,\sigma_{ij}^2)$ with $\sigma_{ij}^2 = \mathrm{Var}(\phi_i(X)\phi_j(X))$.

The term 
$$ 
    n^{-1} \boldsymbol{\varepsilon}_{\theta_0}^\top \phi_i(\boldsymbol{x}) \overset{p}{\longrightarrow} \begin{cases}
        0, & \text{if null hypothesis holds}, \\
        \mathbb{E}(\varepsilon_{\theta_0} \phi_i(X)) \neq 0, & \text{if alternative hypothesis holds}.
    \end{cases}
$$
Putting these results together, we have proved the theorem. 

\subsection{Proof of Corollary \ref{cor:null_distribution}}
\label{sec:proof_null_distribution}
Under the null hypothesis, Theorem \ref{thm:residual_asymptotics} states that for each $j$,
$$ 
    (\sqrt{n})^{-1} \boldsymbol{\varepsilon}_{\theta_0}^\top (\hat{e}_j - e_j)(\boldsymbol{x}) = \sqrt{\lambda_j} \sum_{i \neq j} \frac{\lambda_i}{\lambda_j - \lambda_i} o_p(1).
$$
We only need to show that the infinite summation of these $o_p(1)$ terms is still $o_p(1)$. Note that 
$$ 
    \sum_{j=1}^\infty \sqrt{\lambda_j} \left( \sum_{i \neq j} \frac{\lambda_i}{\lambda_j - \lambda_i} \right) < \sum_{j=1}^\infty \sqrt{\lambda_j} \sum_{i=1} \lambda_i/(\eta) < \infty,
$$
by the compactness of $L_k$ and $L_k^{1/2}$, and $\eta = \inf_{i\neq j \in \mathbb{N}} \{|\lambda_i - \lambda_j|\} >0$.

\subsection{Proof of Lemma \ref{lem:projection_finite_sample}}
\label{sec:proof_projection_finite_sample}
Note that:
\begin{align*}
    \frac{1}{\sqrt{n}}  (\hat{\boldsymbol{\Pi}} \boldsymbol{\varepsilon}_{\hat \theta})^\top \hat{e}_j(\boldsymbol{x}) & = \frac{1}{\sqrt{n}}  \left(  \hat{\boldsymbol{\Pi}} \boldsymbol{\varepsilon_0} +   \hat{\boldsymbol{\Pi}} (\nabla_{\theta} \boldsymbol{\varepsilon}(\theta)|_{\theta = \bar{\theta}})^\top (\hat \theta-\theta_0)  \right)^\top \hat{e}_j(\boldsymbol{x}) \\
    & = \frac{1}{\sqrt{n}}  \left(  \hat{\boldsymbol{\Pi}} \boldsymbol{\varepsilon_0} +   \hat{\boldsymbol{\Pi}} (\nabla_{\theta} \boldsymbol{\varepsilon}(\theta)|_{\theta = \hat{\theta}})^\top (\hat \theta-\theta_0) + \hat{\boldsymbol{\Pi}} O_p(n^{-2\beta}) \right)^\top \hat{e}_j(\boldsymbol{x})\\
    & = \frac{1}{\sqrt{n}} \left( \hat{\boldsymbol{\Pi}} \boldsymbol{\varepsilon_0} \right)^\top \hat{e}_j(\boldsymbol{x}) + o_p(1)
\end{align*}
The first equality comes from the mean value theorem, and the last equality is the consequence of the orthogonality between the matrix $\hat{\boldsymbol{\Pi}}$ and the matrix $\nabla_{\theta} \boldsymbol{\varepsilon}(\theta)|_{\theta = \hat{\theta}} = \mathbb{G}^\top$. $o_p(1)$ is a consequence of $\beta>1/4$

\subsection{Proof of Theorem \ref{thm:bootstrap}}
\label{sec:proof_bootstrap}
\begin{align*}
    \frac{1}{\sqrt{n}} (\boldsymbol{\varepsilon}_{\hat \theta}\odot \boldsymbol{V})^\top (\hat{\boldsymbol{\Pi}})^\top \hat{e}_j(\boldsymbol{x}) & = \frac{1}{\sqrt{n}} (\boldsymbol{\varepsilon}_{\hat \theta}\odot \boldsymbol{V})^\top (\hat{\boldsymbol{\Pi}})^\top \left(e_j(\boldsymbol{x})+ (\hat{e}_j - e_j)(\boldsymbol{x})\right) \\
    & = \frac{1}{\sqrt{n}} (\boldsymbol{\varepsilon}_{\hat \theta}\odot \boldsymbol{V})^\top (\hat{\boldsymbol{\Pi}})^\top e_j(\boldsymbol{x}) + \frac{1}{\sqrt{n}} (\boldsymbol{\varepsilon}_{\hat \theta}\odot \boldsymbol{V})^\top (\hat{\boldsymbol{\Pi}})^\top (\hat{e}_j - e_j)(\boldsymbol{x})\\
    & = \frac{1}{\sqrt{n}} \sqrt{\lambda_j} (\boldsymbol{\varepsilon}_{\theta_0}\odot \boldsymbol{V})^\top (\hat{\boldsymbol{\Pi}})^\top \phi_j(\boldsymbol{x}) + \frac{1}{\sqrt{n}} (\boldsymbol{\varepsilon}_{\theta_0}\odot \boldsymbol{V})^\top (\hat{\boldsymbol{\Pi}})^\top (\hat{e}_j - e_j)(\boldsymbol{x}),
\end{align*}
where the last equality comes from the fact that
\begin{align*}
    \hat{\boldsymbol{\Pi}} (\boldsymbol{\varepsilon}_{\hat \theta}\odot \boldsymbol{V})& = \boldsymbol{\varepsilon}_{\hat \theta}\odot \boldsymbol{V} - \mathbb{G}(\mathbb{G}^\top \mathbb{G})^{-1} \mathbb{G}^\top (\boldsymbol{\varepsilon}_{\hat \theta}\odot \boldsymbol{V}) \\
    & = \boldsymbol{\varepsilon}_{\theta_0}\odot \boldsymbol{V} + \mathbb{G}(\mathbb{G}^\top \mathbb{G})^{-1} \mathbb{G}^\top (\boldsymbol{\varepsilon}_{\theta_0}\odot \boldsymbol{V}) \\
    & = \hat{\boldsymbol{\Pi}} (\boldsymbol{\varepsilon}_{\theta_0}\odot \boldsymbol{V}) \in \mathbb{R}^n.
\end{align*}
By the same argument in Theorem \ref{thm:residual_asymptotics}, we have 
\begin{align*}
    \frac{1}{\sqrt{n}} (\boldsymbol{\varepsilon}_{\theta_0}\odot \boldsymbol{V})^\top (\hat{\boldsymbol{\Pi}})^\top (\hat{e}_j - e_j)(\boldsymbol{x}) = o_p(1),
\end{align*}
since $\mathbb{E}(V_1)=0$, $\mathrm{Var}(V_1)=1$, and is independent of the data. Furthermore, by the multiplier central limit theorem (see \cite{vaart1997weak}), conditional on $\{y_\alpha,x_\alpha\}_{\alpha=1}^n$, we have
$$ 
    \frac{1}{\sqrt{n}} \sqrt{\lambda_j} (\boldsymbol{\varepsilon}_{\theta_0}\odot \boldsymbol{V})^\top (\hat{\boldsymbol{\Pi}})^\top \phi_j(\boldsymbol{x}) \overset{d^*}{\longrightarrow} \sqrt{\lambda_j} W_j,
$$
and the covariance structure is $\mathrm{Var}(W_i,W_j)=\mathbb{E}(\varepsilon_{\theta}^2 \phi_i(X)\phi_j(X)), \, i\neq j$. The rest of the proof is straightforward.

\end{document}